\documentclass[%
 reprint,
superscriptaddress,
 amsmath,amssymb,
 aps,
]{revtex4-2}

\usepackage{graphicx}
\usepackage{dcolumn}
\usepackage{bm}
\usepackage{color,soul}
\usepackage{subcaption}

\begin{document}

\preprint{APS/123-QED}

\title{Generalization of Kirchhoff’s Law of Thermal Radiation: \\The Inherent Relations Between Quantum Efficiency and Emissivity}

\author{M. Kurtulik}
 \altaffiliation[]{Equal contribution as first author.}
 \affiliation{%
Russell Berrie Nanotechnology Institute, Technion -- Israel Institute of Technology. Haifa 3200003, Israel}
 
 \author{M. Shimanovic}
\altaffiliation[]{Equal contribution as first author.}
\affiliation{Department of Mechanical Engineering, Technion -- Israel Institute of Technology. Haifa 3200003, Israel
}

 \author{T. Bar Lev}
\altaffiliation[]{Equal contribution as first author.}
\affiliation{Department of Mechanical Engineering, Technion -- Israel Institute of Technology. Haifa 3200003, Israel
}

 \author{R. Weill}
 \affiliation{Department of Mechanical Engineering, Technion -- Israel Institute of Technology. Haifa 3200003, Israel
}
 \author{A. Manor}
 \affiliation{%
Russell Berrie Nanotechnology Institute, Technion -- Israel Institute of Technology. Haifa 3200003, Israel}
 \author{M. Shustov}
 \affiliation{Department of Mechanical Engineering, Technion -- Israel Institute of Technology. Haifa 3200003, Israel
}
\author{C. Rotschild}%
 \email[Corresponding author: ]{carmelr@technion.ac.il}
\affiliation{%
Russell Berrie Nanotechnology Institute, Technion -- Israel Institute of Technology. Haifa 3200003, Israel}
\affiliation{Department of Mechanical Engineering, Technion -- Israel Institute of Technology. Haifa 3200003, Israel
}




\date{\today}

\begin{abstract}
Planck’s law of thermal radiation depends on the temperature, $T$, and the emissivity, $\epsilon$, of a body, where emissivity is the coupling of heat to radiation that depends on both phonon-electron nonradiative interactions and electron-photon radiative interactions.
Another property of a body is absorptivity, $\alpha$, which only depends on the electron-photon radiative interactions. At thermodynamic equilibrium, nonradiative interactions are balanced, resulting in Kirchhoff’s law of thermal radiation that equals these two properties, i.e., $\epsilon = \alpha$. 
For non-equilibrium, quantum efficiency ($QE$) describes the statistics of photon emission, which like emissivity depends on both radiative and nonradiative interactions. Past generalized Planck’s equation extends Kirchhoff’s law out of equilibrium by scaling the emissivity with the pump-dependent chemical-potential $\mu$, obscuring the relations between the body properties. 
Here we theoretically and experimentally demonstrate a prime equation relating these properties in the form of $\epsilon = \alpha(1-QE)$, which is in agreement with a recent universal modal radiation law for all thermal emitters. 
At equilibrium, these relations are reduced to Kirchhoff’s law. 
Our work was done on photoluminescence. Future expansion to electro-luminescence and other quantum emission processes, will uncover the shared fundamental principles of radiation.

\end{abstract}

\maketitle


\section{\label{sec:level1}Introduction}

Planck thermal radiation is the emission of a body at non-zero temperature. It is characterized by the temperature of the body $T$, and its ability to thermally emit radiation, given by emissivity $\epsilon$, which is a material and cavity property reflecting the contribution of both radiative $\gamma_r$, and non-radiative $\gamma_{nr}$ rates. While $\gamma_{nr}$ couples heat to excite electrons, $\gamma_{r}$ couples the excited electrons to photon emission. In the reverse process, these rates respectively govern non-radiative recombination and photon absorption, denoted by $\alpha$. Although absorptivity depends only on the radiative rates, while emissivity depends on both radiative and non-radiative rates, at thermodynamic equilibrium, thermal excitation is balanced by the non-radiative recombination, leading to the equality between emissivity and absorptivity:
\begin{equation}
    \label{eq. Kirchhoff's law equilibrium}
    \epsilon = \alpha
\end{equation} 
known as Kirchhoff’s law of thermal radiation. Out of equilibrium, i.e., a body emits radiation into an environment having a different temperature or chemical potential, Kirchhoff’s law does not apply, since non-radiative processes are not canceled out (thermal excitation $\neq$ non-radiative recombination). This phenomenon can be seen in semiconductors, light-emitting diodes, atoms and photoluminescence~\cite{Yablonovitch1995}. An extreme example of non-equilibrium is when absorptivity becomes negative at high excitation rates while the emissivity remains positive \cite{miller2012strong,sheng2015device,tran2014high}.

Photoluminescence (PL), as an example of non-equilibrium emission, involves the absorption of photons followed by fast thermalization of the excited electrons and emission of (typically low-energy, red-shifted) photons \cite{stokes1852xxx,vij1998thermoluminescence,liu2006spectroscopic,gaft2015modern,hanninen2011lanthanide,thirumalai2016luminescence,bergman2011handbook}. Since thermalization results in the energy difference between the incoming and outgoing photon energy, it is defined as out of equilibrium. Nevertheless, thermodynamics statistically analyzes PL at quasi-equilibrium
\cite{einstein1916emission,kirchhoff1859uber,Stefan:1879txg,wien1896ueber}, using the conventional thermodynamic variables such as temperature, emissivity, and chemical potential $\mu$ 
\cite{landau1946thermodynamics,wurfel1982chemical}, where the latter describes the excitation above thermal. The generalized Planck’s law describes such non-equilibrium emission
\cite{wurfel1982chemical,ross1967some} as:
\begin{equation} \label{eq. Generalized Planck}
L\left( h\nu,T,\mu \right) = \epsilon(h \nu) \cdot \frac{2h\nu^3}{c^2} \frac{1}{e^{\frac{h\nu-\mu}{k_BT}}-1} \approx L_{Th} e^{\frac{\mu}{k_BT}}
\end{equation}

where L is the spectral radiance (having units of Watts per frequency, per solid angle, and per unit area); $\epsilon$ is the emissivity, which is a material and cavity property that depends on the density of states (DoS) of the radiative and non-radiative transitions; $T$ is the temperature; $h\nu$ is the photon energy; $k_B$ is Boltzmann’s constant; and $L_{Th}$ is the thermal radiance when the pump is off. The chemical potential $\mu$ is the Gibbs free energy per emitted photon and is also the gap that is opened in semiconductors between the quasi-Fermi levels under excitation. Eq.~(\ref{eq. Generalized Planck}) is relevant at a specific frequency band, where the chemical potential is a constant. Without knowing the inherent relation of $\mu(T)$, however, Eq.~(\ref{eq. Generalized Planck}) can only be fitted to an observed PL, without predicting its evolution.
Previous studies on non-equilibrium emission by external pump excitation assume the temperature of the surroundings to be the same as the PL body~\cite{Yablonovitch1995,miller2012strong,sheng2015device,tran2014high}. In such cases, Eq.~(\ref{eq. Generalized Planck}) can be seen as a generalization of Kirchhoff’s law, demonstrating that the line shape of the non-thermal emission is the same as the thermal emission enhanced by the (scalar) chemical potential, $\epsilon(h\nu) e^{\frac{\mu}{k_BT}}$~\cite{kirchhoff1859uber}. 

Under pump excitation, at low temperature, Eq.~(\ref{eq. Generalized Planck}) also defines the PL quantum efficiency, as the ratio between the emitted and absorbed photon rates:
\begin{equation}
QE = \frac{\text{\# of emitted photons}}{\text{\# of absorbed photons}} = \frac{L(T \rightarrow 0,\mu)}{\alpha \cdot L_{pump}(T_p )}
\end{equation}
where $L(T\rightarrow 0,\mu)$ is the rate of photons emitted by a pumped body at $T\rightarrow 0K$, and $L_{pump} (T_p )$ is the incoming photon rate into the system with a brightness temperature $T_p$, which defines the radiance of the pump, at a specific wavelength, as equal to black-body radiance at the temperature $T_p$. $QE$ is a material and geometric property of the body defined at low temperatures where thermal excitation is negligible and similarly to emissivity, it reflects the competition between radiative and nonradiative rates, where higher $\gamma_{r}$ and lower $\gamma_{nr}$ support higher $QE$.

In the formalism of Eq.~(\ref{eq. Generalized Planck}), in contrast to Planck's formula, the emissivity becomes pump-dependent, losing its original meaning as a material and cavity property, and unable to provide insight into the temperature-dependent evolution of the radiation. This state of affairs leaves many open questions. For example, what is the inherent relation between emissivity and absorptivity for a material that thermally emits $(T>0)$ towards an environment at a different temperature? Is there an inherent relation between emissivity and quantum efficiency ($QE$), since both depend on the same parameters? In similarity to Kirchhoff’s law, what is the inherent evolution of non-equilibrium radiation with temperature? The aim of this paper is to answer these questions by extending Kirchhoff’s law to non-equilibrium and to define the inherent relations between fundamental properties of radiation. 

We begin with the rate equation for a 2-level system and demonstrate how spontaneous and stimulated phonon rates lead to the prime equation. Next, we interpret the solution and show how it aligns with previous studies on photoluminescence at elevated temperatures. We continue with a numerical solution for a 3-level system that generalizes the results obtained for the 2-level system. Finally, we complete the study with an experimental demonstration that directly measures emissivity, absorptivity, and quantum efficiency as predicted by the prime equation. We also explain why prior experimental studies did not reveal the inherent relations between these properties.

\section{methods}
\subsection{2-Level System Analysis}
Observing of the rate equations of a 2-level system reveals the inherent summation of quantum and thermal radiations. Following Siegman~\cite{siegman1986lasers}, a photoluminescent body placed in an open cavity is described by the following equations:
\begin{subequations}
\begin{eqnarray}
\label{2-lvl dn_2}
\frac{dn_2}{dt}=&&(n_1-n_2)B_{r12}n_{ph12}-n_2\gamma_{r12}+ 
\\ \nonumber
&&(n_1-n_2)B_{nr12}n_{pn12}-n_2\gamma_{nr12} 
\end{eqnarray}
\begin{eqnarray}
\label{2-lvl dn_ph2}
4\pi \cdot \Delta\nu \cdot \frac{dn_{ph12}}{dt}=&&n_{pump}\Gamma_p-n_{ph12}\Gamma_{12}-
\\ \nonumber
&&(n_1-n_2)B_{r12}n_{ph12}+n_2\gamma_{r12} 
\end{eqnarray}
\end{subequations}
where $n_1,n_2$ are the electron population densities of the ground and excited states ($N=n_1+n_2$);
$\gamma_r,\gamma_{nr}$ are the corresponding radiative and non-radiative spontaneous rates ($\gamma_{r12},\gamma_{nr12}$);
$B_{r12}=\gamma_{r}/DoS_{ph} , B_{nr12}=\gamma_{nr}/DoS_{pn}$ are the Einstein coefficients for the stimulated absorption or emission rates for 
both radiative and non-radiative processes~\cite{einstein1916emission}; $DoS_{ph} , DoS_{pn}$
are the density of states of the photons and phonons, respectively; 
$\Gamma_p , \Gamma_{out}$ are the coupling rates in (via pump) and out of the cavity, respectively; $n_{ph},n_{pn},p_{pump}$ are the field densities for photons, phonons and incoming photons induced by the pump, respectively.
These densities obey $n_{pump}=DoS_{ph} \cdot f(T_p)$, $n_{ph}=DoS_{ph} \cdot f(T)$ and
$n_{pn}=DoS_{pn} \cdot f(T)$, where $f(T) , f(T_p)$ are the distribution functions obeying the Bose-Einstein statistics for black-bodies, according to the temperature, $T$, and brightness temperature of the pump, $T_p$. 
Conventional solutions for the rate equations assume thermal phononic distribution. Our solution is general and considers both spontaneous and stimulated phonon rates. At steady-state, Eq.~(\ref{2-lvl dn_2}) is solved for $n_2$ and substituting it into Eq.~(\ref{2-lvl dn_ph2}) determines the rate of photons leaving the cavity:
\begin{equation}
    n_{ph12}\Gamma_{out} = \Gamma_{out} \frac{n_{pump}\Gamma_p+NB_{nr12}n_{pn12}\frac{\gamma_r}{\gamma_r+\gamma_{nr}}}{\Gamma_{out}+NB_{r12}\left( 
    1-\frac{\gamma_r}{\gamma_r+\gamma_{nr}}
    \right)} \label{photons leaving cavity}
\end{equation}
We distinguish between two mechanisms of emission:
\begin{subequations}
\begin{enumerate}
    \item Quantum emission, which is pump dependent and accounts only for scatterings and PL:
    \begin{equation}
    L_{quantum} = \Gamma_{out} \frac{n_{pump}\Gamma_p}{\Gamma_{out}+NB_{r12}\left( 
    1-\frac{\gamma_r}{\gamma_r+\gamma_{nr}}
    \right)}     \label{R_quantum}    
    \end{equation}
    \item Thermal emission, which is temperature dependent and radiates according to Planck's law of thermal radiation:
    \begin{equation}
    L_{thermal} = \Gamma_{out} \frac{NB_{nr12}n_{pn12}\frac{\gamma_r}{\gamma_r+\gamma_{nr}}}{\Gamma_{out}+NB_{r12}\left( 
    1-\frac{\gamma_r}{\gamma_r+\gamma_{nr}}
    \right)} \label{r_thermal}
    \end{equation}
\end{enumerate}
\end{subequations}

In our model, we assume the fundamental material parameters — $\gamma_r$, $\gamma_{nr}$, $DoS_{ph}$, and $DoS_{pn}$, to be temperature-independent. Consequently, $\epsilon$, $QE$, $B_{r12}$, and $B_{nr12}$ are also temperature-independent. Based on this assumption, the prime equation is derived:
\begin{equation}
 \epsilon = \alpha \times (1-QE)
 \label{eq. prime}
\end{equation}
This equation is reduced to the familiar Eq.~(\ref{eq. Kirchhoff's law equilibrium}) in equilibrium. The complete derivation can be found in Section 3 in the Supplementary Material.

\begin{figure}[h]
\includegraphics[width=\linewidth]{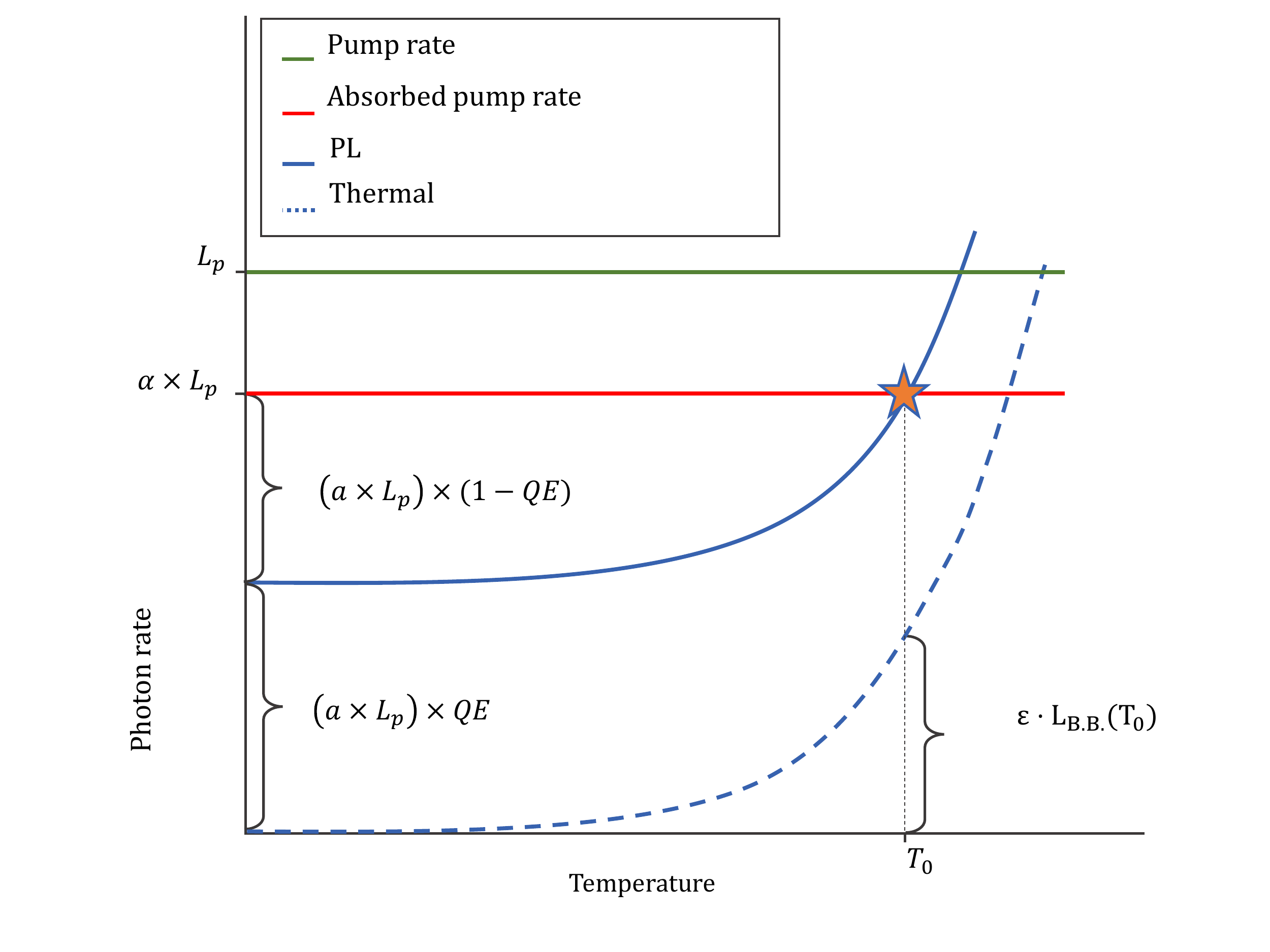}
\caption{\label{fig:1-QE_photon_rate} Illustration of PL vs. temperature.}
\end{figure}

\subsection{Support in Prior Experimental Results}

Recent experimental work on PL at elevated temperatures \cite{manor2015conservation}, shows a quasi-conserved photon emission rate in the low-temperature range, far above thermal emission, which corresponds to the $QE$. At this temperature range, the PL spectrum is blue-shifted with temperature rise. At higher temperatures, the PL rate increases exponentially, approaching the thermal emission. Fig.~\ref{fig:1-QE_photon_rate} illustrates the PL rate evolution with temperature based on~\cite{manor2015conservation}, excluding the very high-temperature range (above 1500K), which will be discussed at the end of the paper. The green line depicts the pump photon rate $L_p$, while the red line depicts the absorbed photon rate $\alpha \times L_p$, where $\alpha$ is the absorptivity. The blue curves are the PL (solid line) and thermal emission (dashed line) for a specific material and geometry for pump-on and pump-off respectively.


At low temperatures, the PL rate is defined by $\alpha \times L_p \times QE$. As the temperature increases, the PL increases monotonically up to a temperature at which the PL photon rate equals the absorbed photon rate (marked with a star in Fig.~\ref{fig:1-QE_photon_rate}), defined as the universal point. The rate of electrons that are photonically excited and recombine non-radiatively at low temperature is $(1-QE) \times\alpha \times L_p$, where at the universal point, this portion is restored by thermal contribution. Therefore, this temperature-dependent evolution equals the thermal emission of a material such that:
\begin{equation}
\label{eq. T0 Tp prime}
\epsilon \times L_{B.B} \left( h\nu,T_0 \right) =  \left(1-QE \right) \times\alpha \times L_p(T_p )
\end{equation}
where $L_{B.B}\left( h\nu,T_0 \right)$ is the black-body emission at $T_0$, which, together with $L_p$, are quantities independent of the specific body properties, while $\epsilon$, $\alpha$ and the $QE$ are material and geometry properties. Thus, we can deduct from this equality the following relation at the universal point:
\begin{equation}
\label{eq. Lbb = Lp}
L_{B.B}(h\nu,T_0) = L_p(T_p )
\end{equation}
By definition, $T_0$ in Eq.~(\ref{eq. Lbb = Lp}) equals the brightness temperature of the pump, $T_p$, manifesting the zeroth law of thermodynamics at the universal point. Inserting Eq.~(\ref{eq. Lbb = Lp}) in Eq.~(\ref{eq. T0 Tp prime}), results in the prime equation, Eq.~(\ref{eq. prime}).
By assuming the material properties to be independent of temperature, the prime equation becomes valid at any temperature.  

Intuitively, according to the prime equation, $\epsilon$ must be reduced, compared to $\alpha$, by the $QE$ amount. This also agrees with the understanding that high-$QE$ materials have low phonon-electron nonradiative rates.

Recent theoretical work on a universal modal radiation law~\cite{miller2017universal} supports a very similar conclusion while analyzing a scattering system in thermodynamic equilibrium. We may treat photoluminescence as part of the  scattering coefficient $|s_p|^2$ and reach our prime equation. The similarity between the two approaches is explained in greater detail in the Supplementary Material in section~1.

\subsection{Agreement With Prior Theoretical Formalism}
The separation between quantum and thermal emission described in~(\ref{R_quantum}),~(\ref{r_thermal}) necessitates the interpretation of the chemical potential $\mu$, in accordance with the generalized Planck's law in Eq.~(\ref{eq. Generalized Planck}), which can be expressed as:
\begin{equation}
    \epsilon \times L_{B.B} e^{\frac{\mu}{k_BT}} = \alpha \times QE \times L_p + \alpha \times (1-QE) \times L_{B.B}
\end{equation}
and by applying Eq.~(\ref{eq. prime}), the chemical potential is reduced to:
\begin{equation}
\begin{aligned}
    \mu &= k_B T \cdot \ln \left( \frac{L_{quantum}}{L_{thermal}} +1 \right) 
    \\ 
    &= k_B T \cdot \ln \left( \frac{L}{L_{thermal}}\right) 
    \label{Chemical potential}  
\end{aligned}
\end{equation}
where $L_{quantum} = \alpha \times QE \times L_p$ and $L_{thermal} = \epsilon \times L_{B.B}$ which is in agreement with~\cite{ross1967some}.

\subsection{3-level system numerical simulation}
Following Siegman~\cite{siegman1986lasers}, we study the case of a 3-level system in a cavity emitting toward a zero-K environment (describing, for example, the 830–900 nm emission lines of $Nd^{+3}$ as depicted in~\cite{manor2015conservation}). Such a detailed balance considers only photonic, electronic, and phononic transitions and omits other processes associated, for example, with defects, which may cause additional temperature-dependent quenching. As such, this model describes the upper limit of temperature-dependent luminescence. Nevertheless, any additional factors can be embedded in the radiative and non-radiative rates for a specific solution. 

\begin{figure*}
     \begin{subfigure}[b]{0.3\textwidth}
         \includegraphics[width=\textwidth]{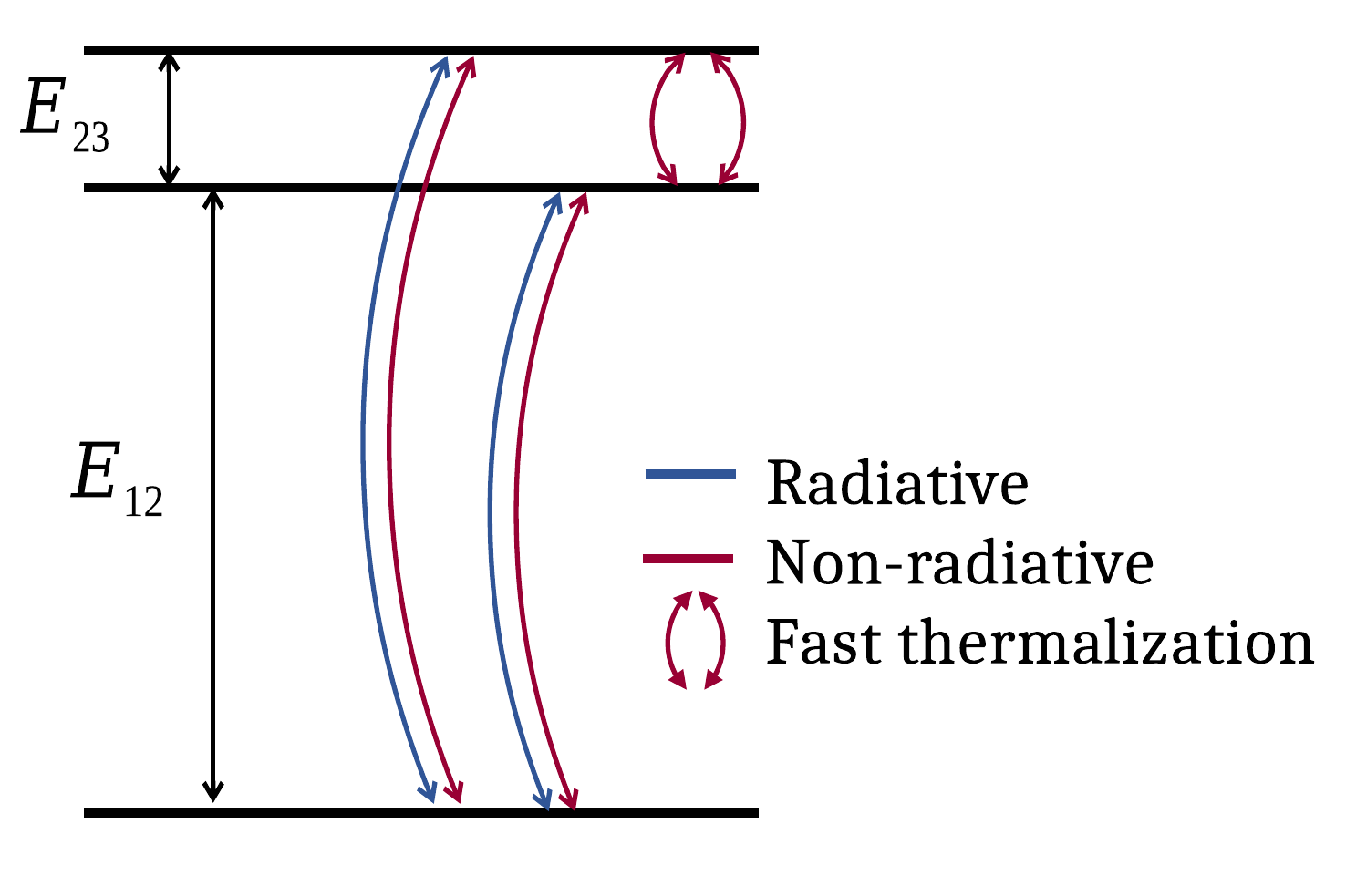}
         \caption{}
         \label{fig:3_level}
     \end{subfigure}
     \hfill
     \begin{subfigure}[b]{0.3\textwidth}
         \includegraphics[width=\textwidth]{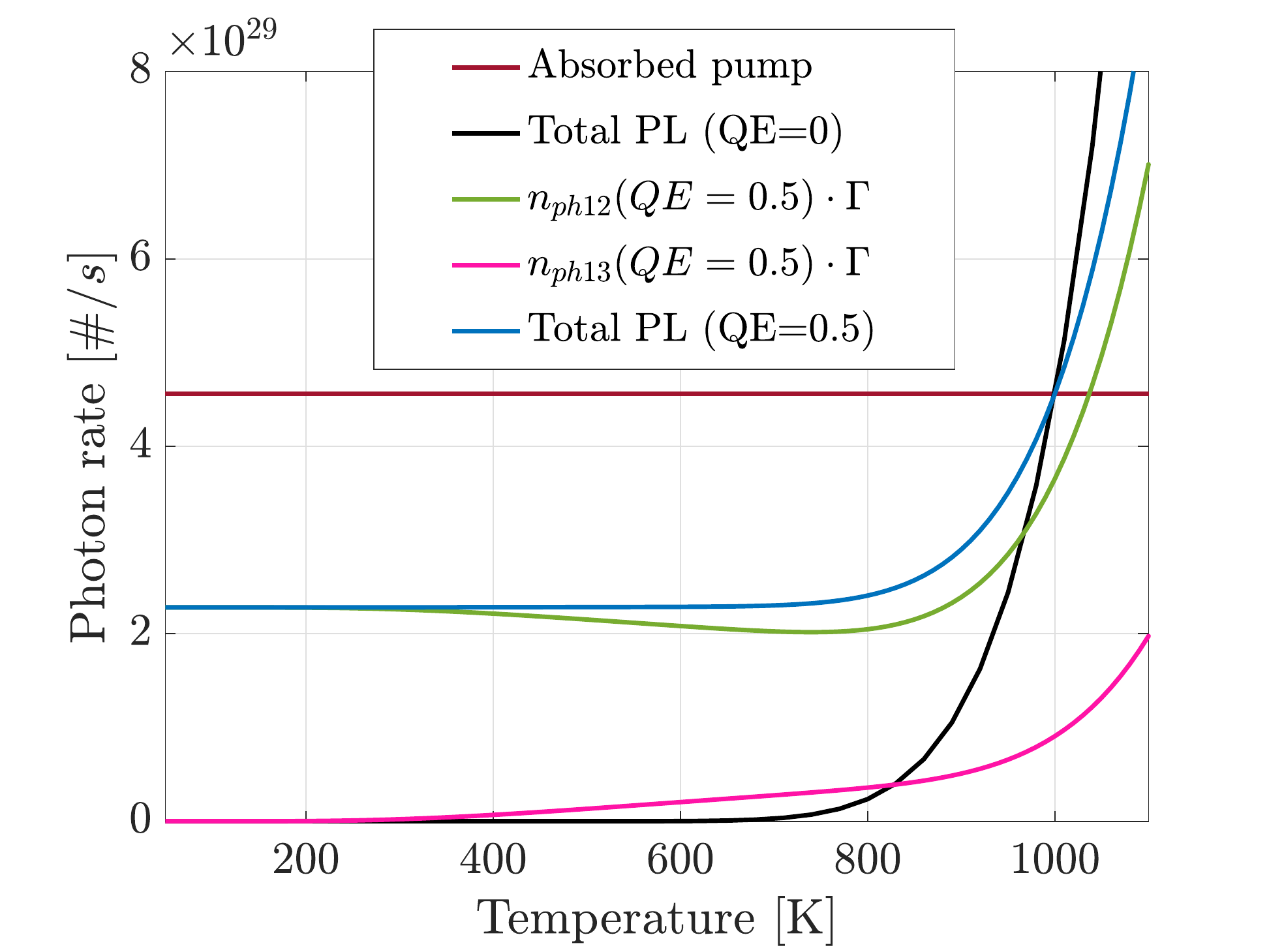}
         \caption{}
         \label{fig:PL_pump}
     \end{subfigure}
     \hfill
     \begin{subfigure}[b]{0.3\textwidth}
         \includegraphics[width=\textwidth]{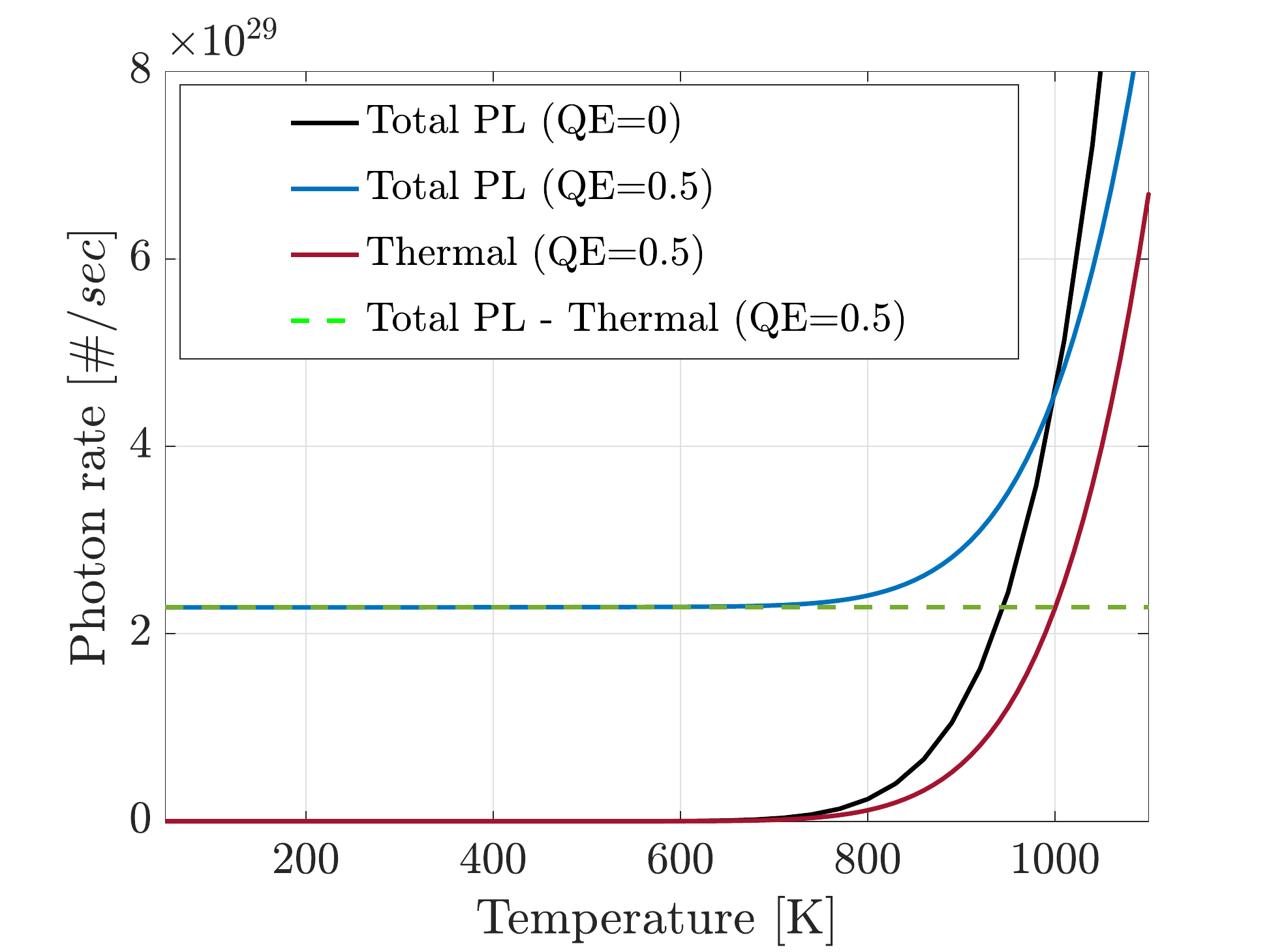}
         \caption{}
         \label{fig:PL_thermal}
     \end{subfigure}
        \caption{A 3-level system with fast non-radiative thermalization between upper energy levels. (b) PL emission for three different QE: $QE=0,\; 0.5$ and 1 systems (black, blue and red lines, respectively), optically excited by a pump at brightness temperature $T_p=1000K$. The green and purple lines show individual emissions from lower and higher excited energy levels for the $QE=0.5$ case. (c) Above the critical temperature $T>T_c$, the PL emission (blue line) is bounded by a thermal body when $QE=0$ (black line) and by the thermal emission for the specific $QE$ ($QE=0.5$, red line). The dashed green line shows the constant difference between the PL emission and the thermal emission for $QE=0.5$.}
\end{figure*}

Fig.~\ref{fig:3_level} shows the considered energy levels having a ground state and a broad excited level consisting of two closely spaced levels, with fast non-radiative thermalization between them $(\gamma_{nr23})$. This drives towards a Boltzmann distribution of the excited electron populations between the two levels, $n_2$ and $n_3$ \cite{wade2000strain} (detailed solution in Section 2 of the Supplementary Material). Also, here, the brightness temperature $T_p$ defines the pump rate within the angular and spectral coupling window $\Gamma_p$. The system is described by:
\begin{subequations}
\begin{eqnarray}
\frac{dn_2}{dt}=&&(n_1-n_2)B_{r12}n_{ph12}-n_2\gamma_{r12}+ 
\\ \nonumber
&&(n_1-n_2)B_{nr12}n_{pn12}-n_2\gamma_{nr12}+
\\ \nonumber
&&(n_3-n_2)B_{nr23}n_{pn23}+n_3\gamma_{nr23} \label{dn2/dt}
\end{eqnarray}
\begin{eqnarray}
\frac{dn_3}{dt}=&&(n_1-n_3)B_{r13}n_{ph13}-n_3\gamma_{r13}+
\\ \nonumber
&&(n_1-n_3)B_{nr13}n_{pn13}-n_3\gamma_{nr13}-
\\ \nonumber
&&(n_3-n_2)B_{nr23}n_{pn23}-n_3\gamma_{nr23} \label{dn3/dt}
\end{eqnarray}
\begin{eqnarray}
4\pi \cdot \Delta\nu \cdot \frac{dn_{ph12}}{dt}=&&n_{pump}\Gamma_p-n_{ph12}\Gamma_{12}-
\\ \nonumber
&&(n_1-n_2)B_{r12}n_{ph12}+n_2\gamma_{r12} \label{dn_ph12/dt}
\end{eqnarray}
\begin{eqnarray}
4\pi \cdot \Delta\nu \cdot \frac{dn_{ph13}}{dt}=&&n_{pump}\Gamma_p-n_{ph13}\Gamma_{13}-
\\ \nonumber
&&(n_1-n_3)B_{r13}n_{ph12}+n_2\gamma_{r13} \label{dn_ph13/dt}
\end{eqnarray}
\end{subequations}
where all the parameters are identical to those introduced in Eq.~(\ref{2-lvl dn_2}) and Eq.~(\ref{2-lvl dn_ph2}), modified to describe a 3-level system.
Solving the equations for different regimes reveals different observables. For simplicity, we plot the solution for various values of $\gamma_{r}$ and $\gamma_{nr}$ as temperature-independent. For the temperature-dependent case, the solution can be depicted as crossing between different curves at different temperatures.
\begin{figure}[t!]
\includegraphics[width=\linewidth]{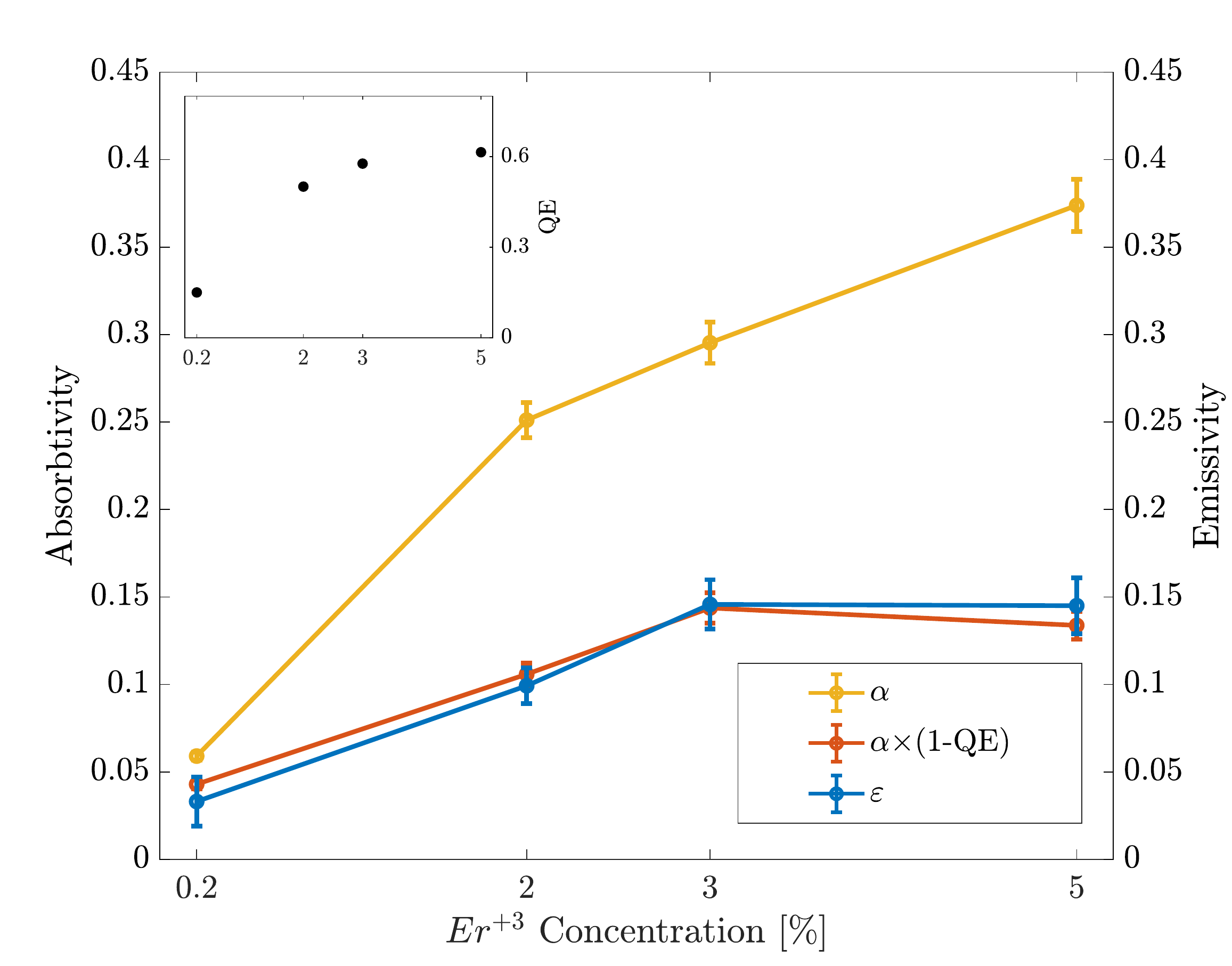}
\caption{\label{fig: abs vs conc}Generalization of Kirchhoff's law for non-equilibrium radiation. Absorptivity of different $QE$ samples, under optical excitation, measured at room temperature (yellow) and multiplied by $(1-QE)$ of each sample (red). The blue line is the measured emissivity of each sample. The inset presents the independently measured $QE$ as a function of Er concentration.}
\end{figure}
\begin{figure}[t!]
\includegraphics[width=\linewidth]{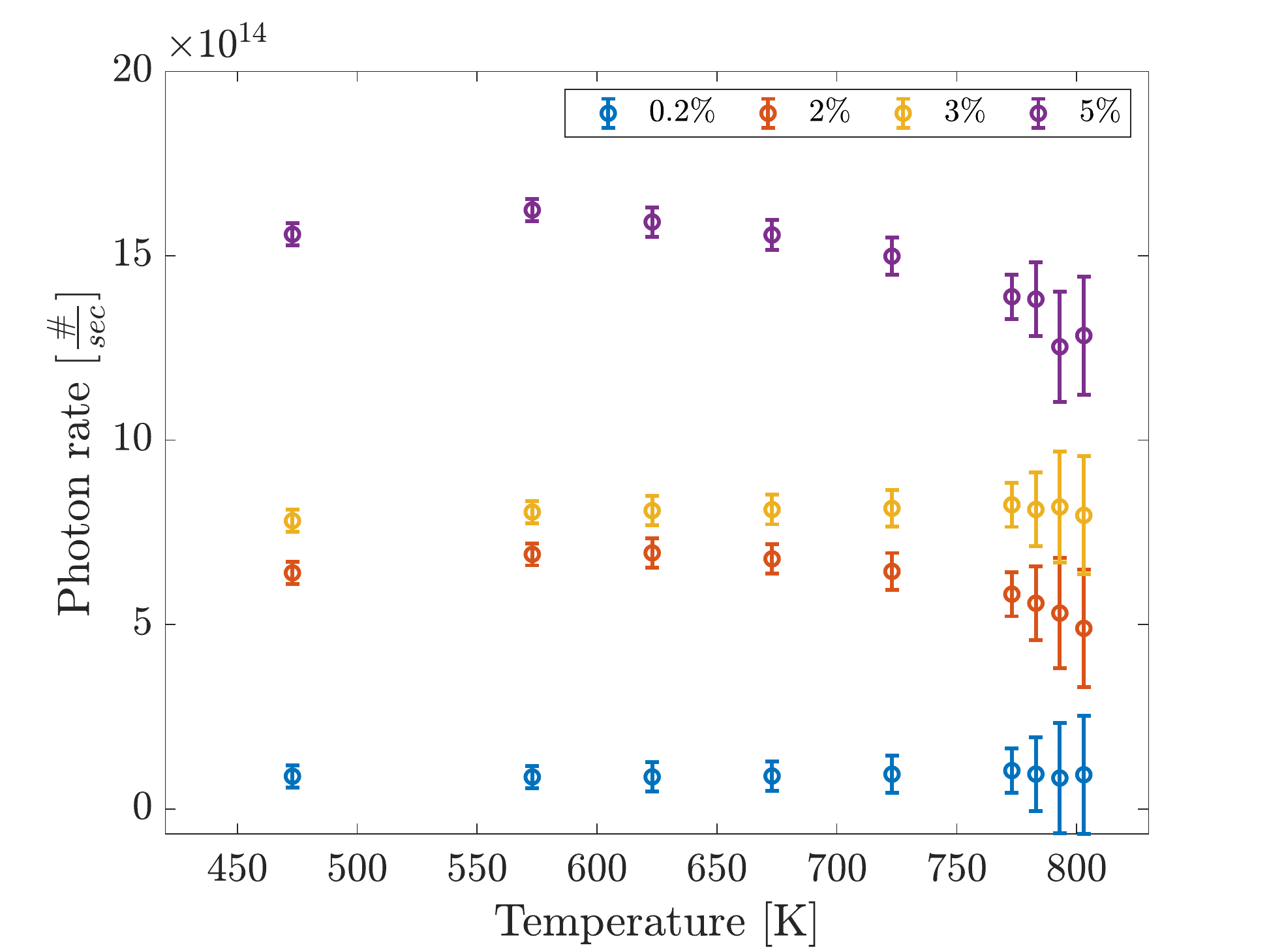}
\caption{\label{Diff_PL_Thermal}Difference between PL and thermal emission rates in the measured temperature range.}
\end{figure}

The solutions for the photon rate (radiative transitions to the ground state) are given by $n_{ph12} \Gamma_{12}$ and $n_{ph13} \Gamma_{13}$, for equally incoming and output coupling rates  
$\Gamma_{12}=\Gamma_{13}=\Gamma_{p}=\Gamma$. 
We also assume an optical excitation $n_{pump12} \Gamma_{p}$ and $n_{pump13} \Gamma_{p}$ at $T_p=1000K$ for various QEs (0, 0.5, and 1). The results are depicted in Fig.~\ref{fig:PL_pump}. These three cases are all chosen to have the same absorptivity, depending on $\gamma_r$, and different losses due to different $\gamma_{nr}$ rates. The red line describes both the absorbed pump rate and the $QE=1$ case. The black line in 
Fig.~\ref{fig:PL_pump} describes the thermal emission, when setting $QE=0$ (while keeping $\gamma_r$ constant), describing the case where $\gamma_{nr12},\gamma_{nr13} \gg \gamma_{r12},\gamma_{r13}$.

Under this regime, nearly all the absorbed photons recombine nonradiatively, which results in thermal emission that increases exponentially until it reaches the $QE=1$ line at the universal point $T=T_p$. This point is the intersection between the three lines ($QE=0$, $QE=1$, and $QE=0.5$). A body with the $QE\cong 0$, emits the same amount of radiation to the environment regardless of the optical excitation. It is in the local equilibrium set by the dominant nonradiative rates, and obeys Kirchhoff’s law. For the PL emission of $QE=0.5$ (The blue line), at $0K$, half of the absorbed photons are lost due to nonradiative recombination $\gamma_{nr}$. Agreeing with the experimental observations~\cite{manor2015conservation}, in the low (non-zero) temperature range, the total photon rate (the sum of $n_{ph12} \Gamma_{12}$ and $n_{ph13} \Gamma_{13}$) is quasi-conserved, accompanied by a blue-shift of the spectrum-emitted photon rate. That is, $n_{ph12} \Gamma_{12}$ decreases while $n_{ph13} \Gamma_{13}$ increases, as the temperature increases. The ratio between these emissions is given by the Boltzmann distribution as long as $\gamma_{nr23} \gg \gamma_{r12}, \gamma_{r13},\gamma_{nr12} ,\gamma_{nr13}$~\cite{wade2000strain}. A further increase in temperature leads to an increase in the photon rate until it reaches the intersection between the $QE=1$ case (red line) and the $QE=0$ case (black line) at the universal point $T=T_p$. The photon rate continues, thereafter, to rise above it.

In the absence of an optical pump ($n_{pump}=0$), one observes the thermal emission for the body for any $QE$ emitting towards a zero-K environment. Fig.~\ref{fig:PL_thermal} shows the PL line for $QE=0.5$ (blue line) in addition to its corresponding thermal emission (red line). For comparison, we also show the thermal emission for the case of $QE=0$ (black line), all have the same $\gamma_r$ values. The difference between the PL and thermal lines for $QE=0.5$ is a constant, invariant of temperature (depicted by the green dashed line). 

This general solution is, as far as we know, the first explanation for the experimentally observed transition from the rate-conservation region accompanied by a blue shift to thermal emission, where the photon rate increases at any wavelength, as shown in \cite{manor2015conservation}. Evidently, the thermal emission for $QE>0$ is reduced, compared to the thermal emission of $QE=0$, due to a lower $\gamma_{nr}$ value compared to $\gamma_r$. In addition, the PL emission beyond the universal point is restricted to remaining between the $QE=0$ line and the thermal line for the same $QE$.

We find from figure Fig.~\ref{fig:PL_thermal} that the ratio between the thermal emission of a body (for any $QE>0$, in the absence of a pump) and the emission curve of a thermal body having the same absorptivity ($QE=0$ and $\epsilon_{(QE=0)}=\alpha$) is a temperature-independent and $QE$-dependent constant we name $\epsilon_{(QE>0)}$, which results in $\epsilon_{(QE>0)} =1-QE$. It is evident, that the thermal contribution leads to recovery of the universal point and balancing nonradiative recombination which are proportional to $\alpha \times (1-QE)$. As suggested by Eq.~(\ref{eq. prime}), the total emissivity can be expressed as:
\begin{equation}
\epsilon = \epsilon_{(QE=0)} \times \epsilon_{(QE>0)} = \alpha \times (1-QE) \label{eq. total emissivity}
\end{equation}
The total thermal emission of a body toward a zero-K environment (in absence of a pump) is: 
\begin{equation}
L\left( h\nu,T \right) = \Gamma \times \alpha \times (1-QE) \times \frac{2h\nu^3}{c^2} \frac{1}{e^{\frac{h\nu}{k_BT}}-1} \label{total thermal emis 0K}
\end{equation}

For the case where $\gamma_{r}$ and $\gamma_{nr}$ are temperature dependent, as for Planck’s radiation, the emission curve crosses from one $QE$ line to the other, yet still reaching the universal point. For this reason, Eq.~(\ref{eq. total emissivity})  is valid also in the form $\epsilon(T) = \alpha(T) \times (1-QE(T))$.

In addition, the equations show that in the case where the coupling rate $\Gamma \to 0$, any material approach $QE=0$. This means that the emission from any material enclosed in an optical cavity becomes black-body emission. The cavity reduces the photons escaping the system thereby effectively reducing $\gamma_{r}$, while unchanging $\gamma_{nr}$ which is a material property. The relation $\gamma_{nr} \gg\ \gamma_{r}$, together with multiple self-absorption events in the cavity that effectively enhance absorptivity results in a black-body.

\section{Experiment}
Conventional absorptivity measurements are conducted in specific directions, while emissivity and $QE$ experiments are performed in an integration sphere, which measures hemispherical radiation. Such measurements do not align with Eq.~(\ref{eq. prime}). Instead, it is necessary to measure hemispherical absorptivity, which is the absorptivity at all directions. This method is elaborated in Section 6 of the Supplementary Materials.

\begin{figure}[t]
    \centering
    \begin{subfigure}{0.4\textwidth}
    \includegraphics[width=\textwidth]{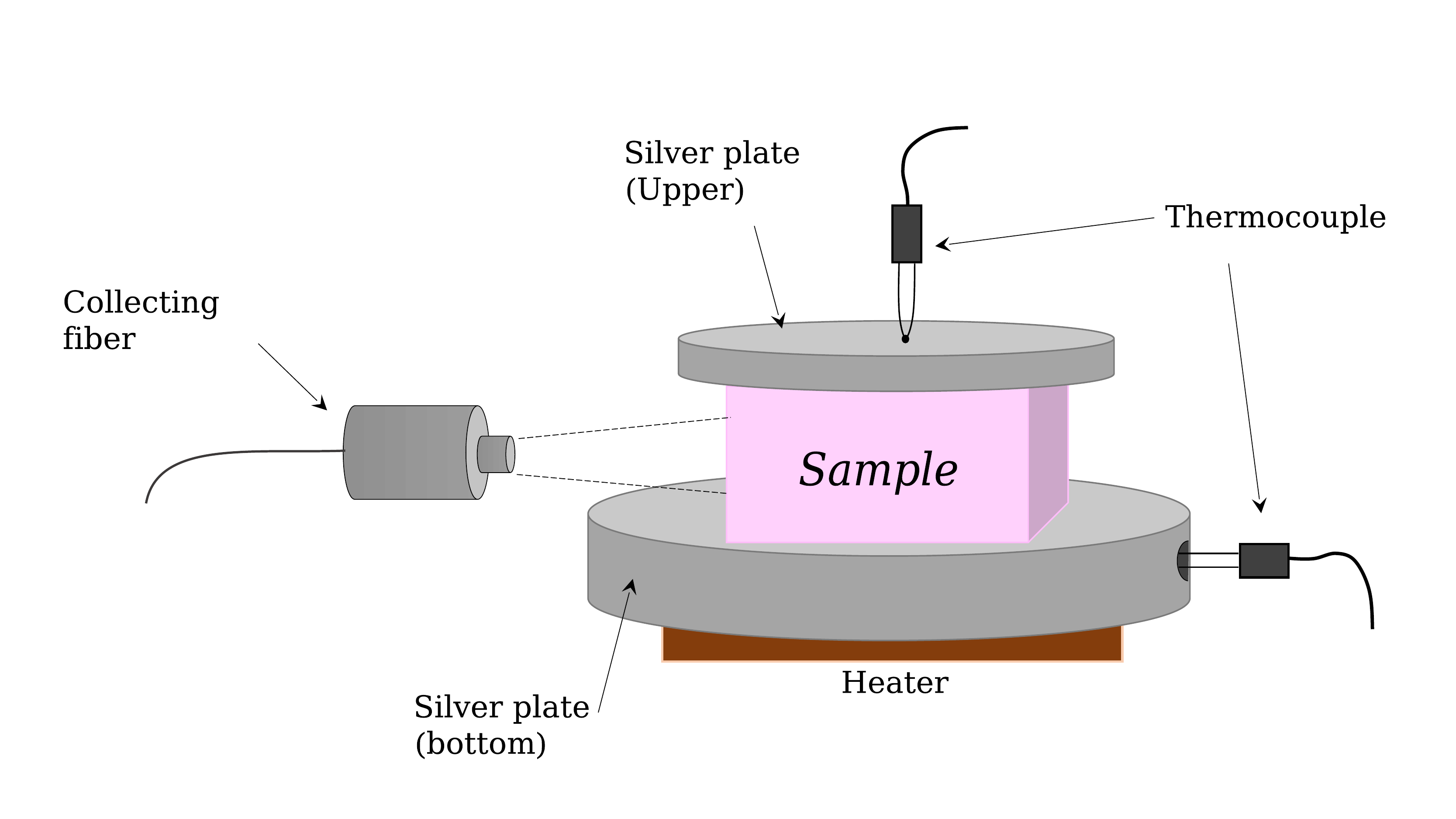}
    \caption{}
    \label{fig: setup}
    \end{subfigure}
    \begin{subfigure}{0.4\textwidth}
    \includegraphics[width=\textwidth]{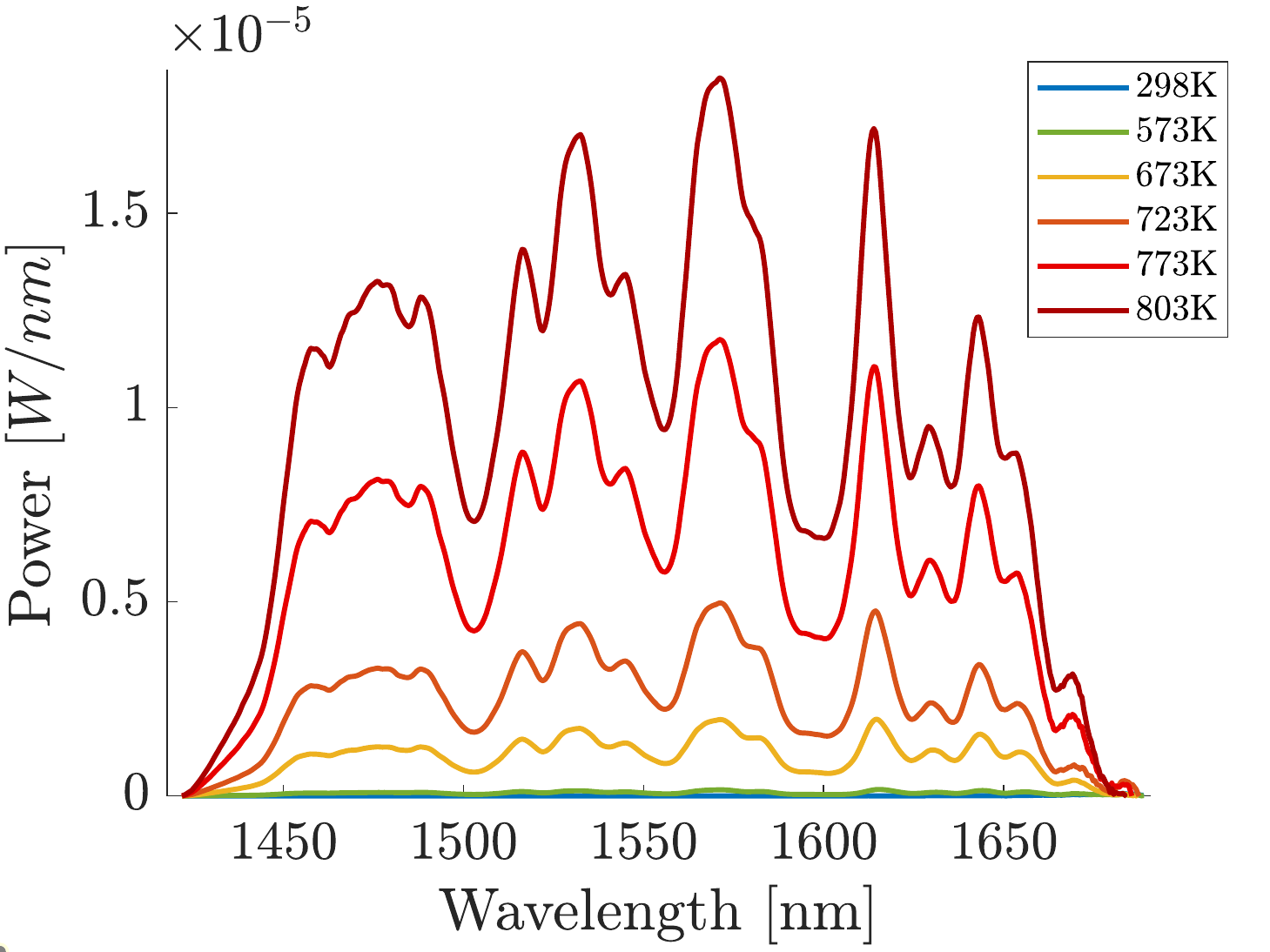}
    \caption{}
    \label{fig: thermal spec 5 percent}
    \end{subfigure}
    \caption{\label{fig: exp setup}Experimental setup and measured power spectra of the thermal emission. (a) The experimental setup. Each Er:YAG sample is placed on a silver plate that is positioned on a heater. The lower silver plate is grooved to align the sample (shown in the inset). The sample is heated by a PID-controlled heater. The temperature of the lower plate is measured by a thermocouple placed in a blind hole, drilled in the silver plate. The emitted thermal radiation of the sample is collected by a fiber and the power spectrum is measured by a calibrated spectrometer. To increase emission into the peripheral directions, the sample is covered on top with a second (non-heated)silver plate. (b) The power spectrum of the thermal emission of the 5\% Er:YAG crystal.}
\end{figure}

For measuring the emissivity of materials with different $QE$s, we choose Erbium-doped YAG (Yttrium Aluminum Garnet, $Y_3 Al_5 O_{12}$) crystals with four different $Er^{+3}$ ion concentrations having mass ratios of 0.2\%, 2\%, 3\% and 5\% \cite{torsello2004origin}. The different concentrations affect the $QE$ through self-quenching, with minimal change in the band gap. The dimensions of the 0.2\% and 3\% $Er^{+3}$ ion concentration crystals are 10×10×4 $mm^3$ (Roditi), while the 2\% and 5\% Er:YAG crystals are 11×11×5 $mm^3$ (Crytur). As illustrated in 
Fig.~\ref{fig: setup}, each sample is placed on a silver plate, which is heated by a ceramic heater (Thorlabs HT19R). Due to the high thermal conductivity of silver, along with its relatively low emissivity, it is possible to uniformly heat the sample, as well as  perform temperature and thermal emission measurements. The temperature of the silver plate is measured with a k-type thermocouple and controlled with a PID controller (Eurotherm 2416). The thermocouple is placed inside a blind hole, with its tips at the center of the silver plate. The sample is covered with an additional silver plate, having an additional thermocouple to monitor and minimize the temperature gradient over the sample. The power spectrum is acquired using multimode fiber, coupled to a calibrated spectrometer (Cornerstore 260) connected to an InGaAs detector (Andor iDus), placed in the vicinity of the sample face. The acquired power spectrum is normalized by the emission area and solid angle. As an example, the thermal emission of the 5\% Er:YAG crystal at the temperature range of 296K-803K is presented in 
Fig.~\ref{fig: thermal spec 5 percent}. The thermal emissions of the other three samples are presented in the Supplementary Material, section 4. A detailed description of the measured emission normalization in the free-space system is given in the Supplementary Material, section 5.

Fig.~\ref{fig: abs vs conc} depicts the obtained emissivity and absorptivity, proving experimentally the generalization of Kirchhoff’s law as presented by Eq.~(\ref{eq. prime}). We note that the measured emissivity, after normalization is the hemispherical emissivity. This value is compared to the hemispherical absorptivity measured inside an integration sphere (see Supplementary Material, section 6). The yellow line depicts the average hemispherical absorptivity at the spectral range (1450–1650 nm) of Erbium samples having different $QE$s. The $QE$s are measured at room temperature~\cite{de1997improved} and presented in the inset as a function of Erbium concentration. The blue line presents the hemispherical emissivity of each sample measured at the same spectral band as the absorptivity. We present the results obtained for 803K, while in Section 7 of the Supplementary Material we show the emissivity vs. temperature, which in our case is constant. Had the $QE$ of the samples been zero, the blue and yellow lines would coincide, representing $\epsilon=\alpha$. From Fig.~\ref{fig: abs vs conc} it is clear that the measured emissivity (blue line) coincides with the absorptivity multiplied by $(1-QE)$ of each sample (red line).

Finally, we experimentally measure our hypothesis that the PL emission is the sum of a constant ($QE$) contribution at zero-Kelvin and a temperature-dependent thermal emission. Fig.~\ref{Diff_PL_Thermal} shows the measured difference between the PL and emissivity of each sample vs. the temperature at ranges where the thermal emission can be obtained accurately (300K–750K). Clearly, all measurements show a constant $QE$ contribution over thermal emission while both increase (500K–750K). The temperature gradient is noticeable only at high temperatures where it reaches the maximal gradient value of 50K. The error bars represent the resulting uncertainty in the photon rate. We note that in~\cite{manor2015conservation} the PL merges with thermal emission at temperatures above 1600K. This can be interpreted as quenching of the $QE$ at high temperatures or as a result of the uncertainty in temperature that leads to high uncertainty in the flux.
\section{Conclusion}
In this paper, we theoretically and experimentally demonstrate the prime equation for emission out of equilibrium as a material property of the form of $\epsilon=\alpha(1-QE)$, relating emissivity, absorptivity, and $QE$. At equilibrium, this relation is reduced to Kirchhoff’s law due to the balance of non-radiative (thermal) processes. Starting from the theoretical model describing a 2-level system through the detailed balance approach, we derive this relation analytically and simulate it for a 3-level system. Experimentally, we demonstrate the prime equation through direct measurement of its properties. We observe at high $QE$ that the emissivity is three-fold lower than absorptivity, with an excellent fit to the prime equation. 
Our work was focused on photoluminescence. As we look ahead, expanding into areas like electro-luminescence and other quantum emission processes will help uncover the underlying shared principles of radiation.
\begin{description}
\item[Funding]
European Union’s Seventh Framework Program (H2020/2014-2020]) under grant agreement n° 638133-ERC-ThforPV.
\\ \\
Israel Science Foundation (Grant No. 1230/21).
\item[Disclosures]
The authors declare no conflicts of interest.
\item[Supplemental document]
See Supplementary Material for supporting content.
\end{description}

\providecommand{\noopsort}[1]{}\providecommand{\singleletter}[1]{#1}%

\end{document}


\preprint{APS/123-QED}

\title{Supplementary Material for "Generalization of Kirchhoff’s Law of Thermal Radiation: The Inherent Relations Between Quantum Efficiency and Emissivity"}

\author{M. Kurtulik}
 \altaffiliation[]{Equal contribution as first author.}
 \affiliation{%
Russell Berrie Nanotechnology Institute, Technion -- Israel Institute of Technology. Haifa 3200003, Israel}
 
 \author{M. Shimanovic}
\altaffiliation[]{Equal contribution as first author.}
\affiliation{Department of Mechanical Engineering, Technion -- Israel Institute of Technology. Haifa 3200003, Israel
}

 \author{T. Bar Lev}
\altaffiliation[]{Equal contribution as first author.}
\affiliation{Department of Mechanical Engineering, Technion -- Israel Institute of Technology. Haifa 3200003, Israel
}

 \author{R. Weill}
 \affiliation{Department of Mechanical Engineering, Technion -- Israel Institute of Technology. Haifa 3200003, Israel
}
 \author{A. Manor}
 \affiliation{%
Russell Berrie Nanotechnology Institute, Technion -- Israel Institute of Technology. Haifa 3200003, Israel}
 \author{M. Shustov}
 \affiliation{Department of Mechanical Engineering, Technion -- Israel Institute of Technology. Haifa 3200003, Israel
}
\author{C. Rotschild}%
 \email[Corresponding author: ]{carmelr@technion.ac.il}
\affiliation{%
Russell Berrie Nanotechnology Institute, Technion -- Israel Institute of Technology. Haifa 3200003, Israel}
\affiliation{Department of Mechanical Engineering, Technion -- Israel Institute of Technology. Haifa 3200003, Israel
}




\date{\today}

\maketitle
\renewcommand{\theequation}{S.\arabic{equation}}
\renewcommand{\thefigure}{S.\arabic{figure}}

\section{Explanation of the agreement between the generalized Kirchhoff's law of thermal radiation and the scattering model}

In the Generalized Kirchhoff’s law of thermal radiation, absorptivity describes the rate of excited electrons by a light source and is defined as $\alpha=1-\rho-\tau$, where $\rho$ and $\tau$ are the reflection and transmission coefficients, respectively. These electrons recombine radiatively or non-radiatively (defined by $QE$). The photoluminescence rate normalized by the pump, $PL_n$, and absorptivity are related as follows:
\begin{equation*}
PL_n=\alpha \times QE
\end{equation*}
Following Eq.~(8) from~\cite{miller2017universal}, we generalize the scattering coefficient $|s_p|^2$ as ``everything that is exiting the material'' in normalized units (including reflection, transmission, and $PL_n$):

\begin{equation} \label{eq. scattering}
|s_p|^2=\rho+\tau+\alpha \times QE = 1- \alpha + \alpha \times QE
\end{equation}
Following Eq.~(9) from~\cite{miller2017universal}, for the absorptivity of the system and Eq.~(12) from~\cite{miller2017universal}, substituting absorptivity and emissivity according to Kirchhoff’s law in equilibrium, we get:
\begin{equation*}
\epsilon_M{}_p = \alpha_M{}_p = 1 - |s_p|^2 
\end{equation*}
Using~\eqref{eq. scattering}, this results in:
\begin{equation*}
\epsilon_M{}_p = 1-(1 - \alpha +\alpha \times QE) = \alpha \times (1-QE)
\end{equation*}
In addition, Eq.~(10) from~\cite{miller2017universal} defines the linear superposition between scattered light and thermal emission. We show in our model the same linear superposition between PL (at 0K) and thermal emission.

\section{Boltzmann distribution of excited electron populations between two energy levels}
The electron population in the three-level systems is described by Eq.~(12a) and~(12b) in the paper. For convenience, we present these equations here as well:
\begin{subequations}
\label{eq:whole}
\begin{eqnarray}
\frac{dn_2}{dt}=&&(n_1-n_2)B_{r12}n_{ph12}-n_2\gamma_{r12}+ 
\label{eq dn2}
(n_1-n_2)B_{nr12}n_{pn12}-n_2\gamma_{nr12}+
\\ \nonumber
&&(n_3-n_2)B_{nr23}n_{pn23}+n_3\gamma_{nr23}
\end{eqnarray}
\begin{eqnarray}
\frac{dn_3}{dt}=&&(n_1-n_3)B_{r13}n_{ph13}-n_3\gamma_{r13}+ (n_1- \label{eq dn3}
n_3)B_{nr13}n_{pn13}-n_3\gamma_{nr13}-
\\ \nonumber
&&(n_3-n_2)B_{nr23}n_{pn23}-n_3\gamma_{nr23}
\end{eqnarray}
\end{subequations}
The last two terms in both equations are the non-radiative interactions between $n_2$ and $n_3$. $\gamma_{nr23}$ is the spontaneous non-radiative rate and $B_{nr23}=\gamma_{nr23}/DoS_{pn}$ is the stimulated non-radiative rate. The electronic levels interact with the phonon field, given by the equilibrium distribution $n_{pn23} = DoS_{pn} \left[ exp\left(\frac{E_{23}}{k_BT} \right)-1 \right]^{-1}$. Thus, we can rewrite these two last terms from Eq.~\eqref{eq dn2} as:

\begin{equation}
\begin{aligned}
(n_3-n_2)B_{nr23}n_{pn23}-n_3\gamma_{nr23} &&= \;\; &\gamma_{nr23}\left( \left[ exp\left(\frac{E_{23}}{k_BT} \right)-1 \right]^{-1} \left[ n_3-n_2 \right]+n_3 \right) 
\\
&&= \;\;&n_3\gamma_{nr23}\left( \left[ exp\left(\frac{E_{23}}{k_BT} \right)-1 \right]^{-1} \left[ 1-\frac{n_2}{n_3} \right]+1 \right)
\\
&&= \;\;&n_3\gamma_{nr23}\left( 1- \frac{\frac{n_2}{n_3}-1}{ exp\left(\frac{E_{23}}{k_BT} \right)-1}  \right)  
\end{aligned}
\end{equation} 
\\ \\
For the case of dominant $\gamma_{nr23}$, at steady-state where $\frac{dn_2}{dt} = \frac{dn_3}{dt} =0$, the equality can be satisfied when these terms approach zero, leading to a Boltzmann distribution:
\begin{equation}
\frac{n_2}{n_3} \approx exp\left(\frac{E_{23}}{k_BT} \right)
\end{equation}

\section{Analytical derivation of the generalized Kirchhoff law $\epsilon = \alpha \times (1-QE)$ for a two-level system}
In Eq. (4), (5a) and (5b) in the paper, we derived the theoretical formalism for the populations of excited electrons and photons within a cavity with a two-level system. This formalism enables the calculation of the rate of photon emission from the cavity. We subsequently demonstrated that this emission consists of two mechanisms: quantum and thermal. For convenience, we have rewritten the equations and will now proceed to prove the primary equationn.

The rate of photons leaving the cavity:
\begin{equation}
    n_{ph12}\Gamma_{out} = \Gamma_{out} \frac{n_{pump}\Gamma_p+NB_{nr12}n_{pn12}\frac{\gamma_r}{\gamma_r+\gamma_{nr}}}{\Gamma_{out}+NB_{r12}\left( 
    1-\frac{\gamma_r}{\gamma_r+\gamma_{nr}}
    \right)}
    \label{photons leaving cavity}
\end{equation}
Where:
\begin{subequations}
        \begin{equation}
    L_{quantum} = \Gamma_{out} \frac{n_{pump}\Gamma_p}{\Gamma_{out}+NB_{r12}\left( 
    1-\frac{\gamma_r}{\gamma_r+\gamma_{nr}}
    \right)}     \label{R_quantum}    
    \end{equation}
    \begin{equation}
    L_{thermal} = \Gamma_{out} \frac{NB_{nr12}n_{pn12}\frac{\gamma_r}{\gamma_r+\gamma_{nr}}}{\Gamma_{out}+NB_{r12}\left( 
    1-\frac{\gamma_r}{\gamma_r+\gamma_{nr}}
    \right)} \label{r_thermal}
    \end{equation}
\end{subequations}

We refer to the ratio between the emitted (or reflected) and incoming photons into the cavity as the external quantum efficiency (EQE):
\begin{equation}
    EQE \equiv \frac{n_{ph12}\Gamma_{out}}{n_{pump}\Gamma_p} = \Gamma_{out}\frac{1+\frac{NB_{nr12} n_{pn12}X}{n_{pump}\Gamma_p}}{\Gamma_{out}+NB_{r12}(1-X)} \label{eq. EQE}
\end{equation}
\\ \\
Where $X=\frac{\gamma_r}{\gamma_r + \gamma_{nr}}$. 
\\ \\
For \underline{any} $\gamma_r$, we can examine the case of $QE \rightarrow 0$ where $\gamma_{nr} \gg \gamma_r$ and Eq.~(\ref{eq. EQE}) is reduced to:
\begin{equation}
    EQE_{(QE \rightarrow 0)} = \frac{1}{1+\frac{NB_{r12}}{\Gamma_{out}}}
\end{equation}
EQE can be positive while $QE \rightarrow 0$ when the non-absorbed pump is reflected off the cavity. In this case, absorptivity is:
\begin{equation}
    \alpha = 1- EQE_{(QE \rightarrow 0)} = \frac{1}{1+\frac{\Gamma_{out}}{NB_{r12}}} \label{eq. EQE for QE=0}
\end{equation}
which is independent of $\gamma_{nr}$. At 0K, phonons are not thermally excited ($n_{pn12} \rightarrow 0$) and Eq.~(\ref{eq. EQE}) becomes:
\begin{equation}
    EQE_{(T \rightarrow 0)}=\frac{\Gamma_{out}}{\Gamma_{out}+NB_{r12}(1-X)}
\end{equation}
\\
The quantum efficiency is defined only at 0K, where the thermal emission vanishes, and it equals to the $EQE_{(T \rightarrow 0)}$ minus the photons reflected off the cavity, normalized to the pump absorbed photons. Therefore:
\begin{equation}
\begin{aligned}
QE_{(T \rightarrow 0)} &&\equiv& \frac{n_{ph12}\Gamma_{out}-(\alpha-1) n_{pump} \Gamma_p}{\alpha \cdot n_{pump}\Gamma_p}
\\
&&=& \frac{EQE_{(T \rightarrow 0)}}{\alpha} - \frac{1-\alpha}{\alpha}
\\
&&=& \frac{X}{1+\frac{NB_{r12}}{\Gamma_{out}}(1-X)}
\label{eq. QE at T=0}
\end{aligned}
\end{equation} 
In order to relate the EQE and the absorption coefficient, $\alpha$, we simplified Eq.~(\ref{eq. QE at T=0}):
\begin{equation}
    \alpha(1-QE) = \frac{NB_{r12}(1-X)}{\Gamma_{out}+NB_{r12}(1-X)}
    \label{eq. a(1-QE)}
\end{equation} \\
The rate of photons leaving the cavity when the pump is off is the thermal emission, Eq.~(\ref{r_thermal}). After substituting $n_{pn}$ it can rewritten as:
\begin{equation}
    n_{ph12}\Gamma_{out} =L_{thermal} = \Gamma_{out} f(T) \frac{NB_{nr12}DoS_{pn}X}{\Gamma_{out}+NB_{r12}(1-X)} \label{eq. R_th pump off}
\end{equation}
\\
Considering the relations between the stimulated and spontaneous rates, we get:
\begin{equation}
    B_{nr12} = B_{r12}\frac{1-X}{X} \cdot \frac{DoS_{ph}}{DoS_{pn}} 
    \label{eq.14}
\end{equation}
\\
Substituting back into Eq.~(\ref{eq. R_th pump off}) yields:
\begin{equation}
         L_{thermal} = \Gamma_{out} DoS_{ph} \cdot f(T) \frac{NB_{r12}(1-X)}{\Gamma_{out}+NB_{r12}(1-X)} \label{eq.15}
\end{equation}
\\
Using Eq.~(\ref{eq. a(1-QE)}), the thermal emission becomes:
\begin{equation}
    L_{thermal} = \Gamma_{out} \alpha \left( 1-QE \right) DoS_{ph} \cdot f(T) \label{eq. prime R_th}
\end{equation}
\\
Comparing Eq.~(\ref{eq. prime R_th}) with Plank's law of thermal radiation:
$L_{BB} = \epsilon \cdot DoS_{ph} \cdot f(T)$
, results in the prime equation:
\begin{equation}
    \epsilon = \alpha \left( 1-QE \right)
\end{equation}
\\
In this model, we assumed that the fundamental material parameters 
$\gamma_r , \gamma_{nr} , DoS_{ph} , DoS_{pn}$ 
are temperature independent. Therefore, the emissivity, $\epsilon$, quantum efficiency and stimulated rates 
$B_{r12} , B_{nr12}$
are also temperature independent.
\section{Thermal emission spectra of all measured materials, 0.2\%, 2\% and \%3 Er:YAG}
\begin{figure*}[]
     \centering
     \begin{subfigure}[b]{0.3\textwidth}
         \includegraphics[width=\textwidth]{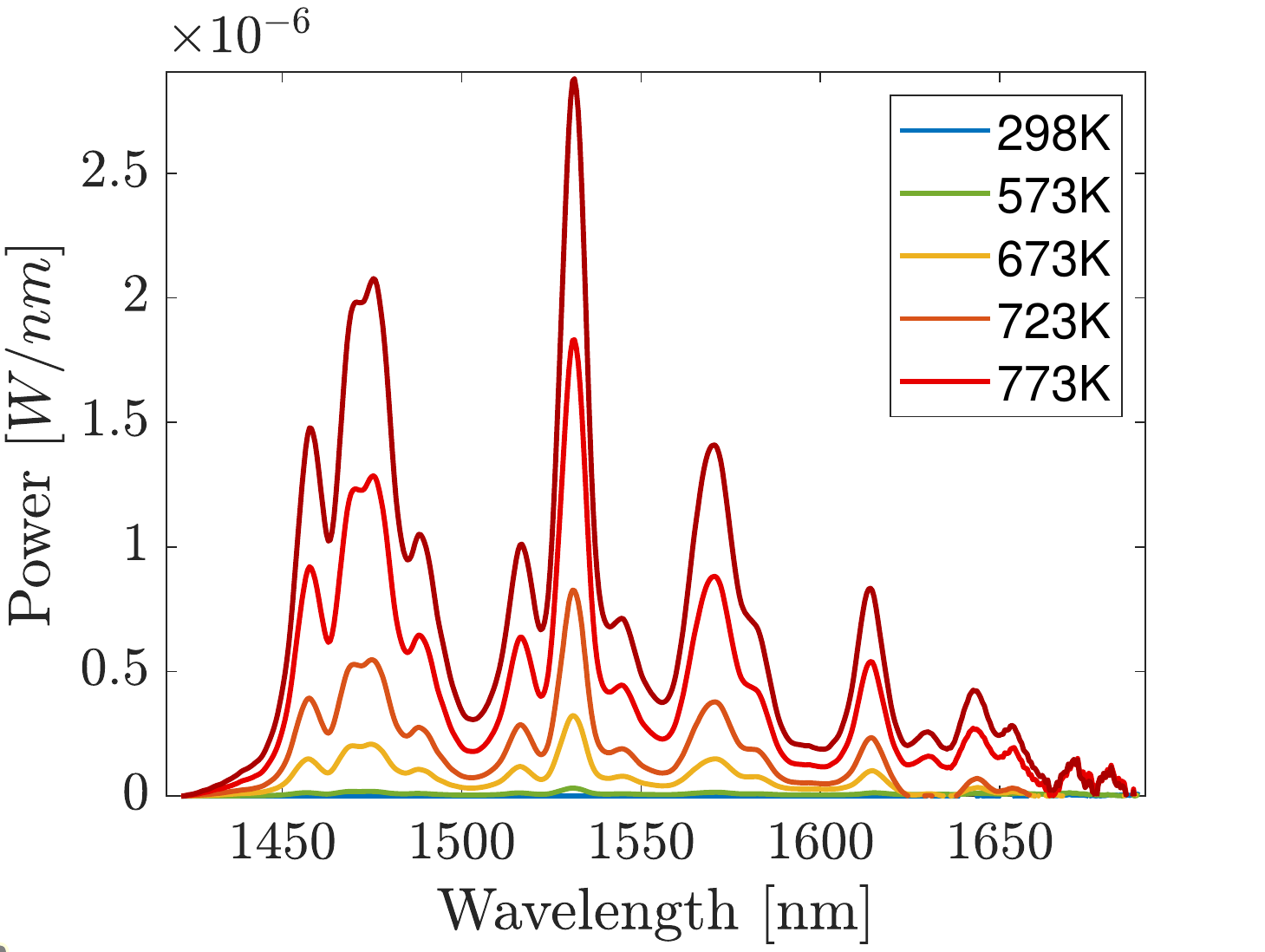}
         \caption{}
     \end{subfigure}
     \hfill
     \begin{subfigure}[b]{0.3\textwidth}
         \includegraphics[width=\textwidth]{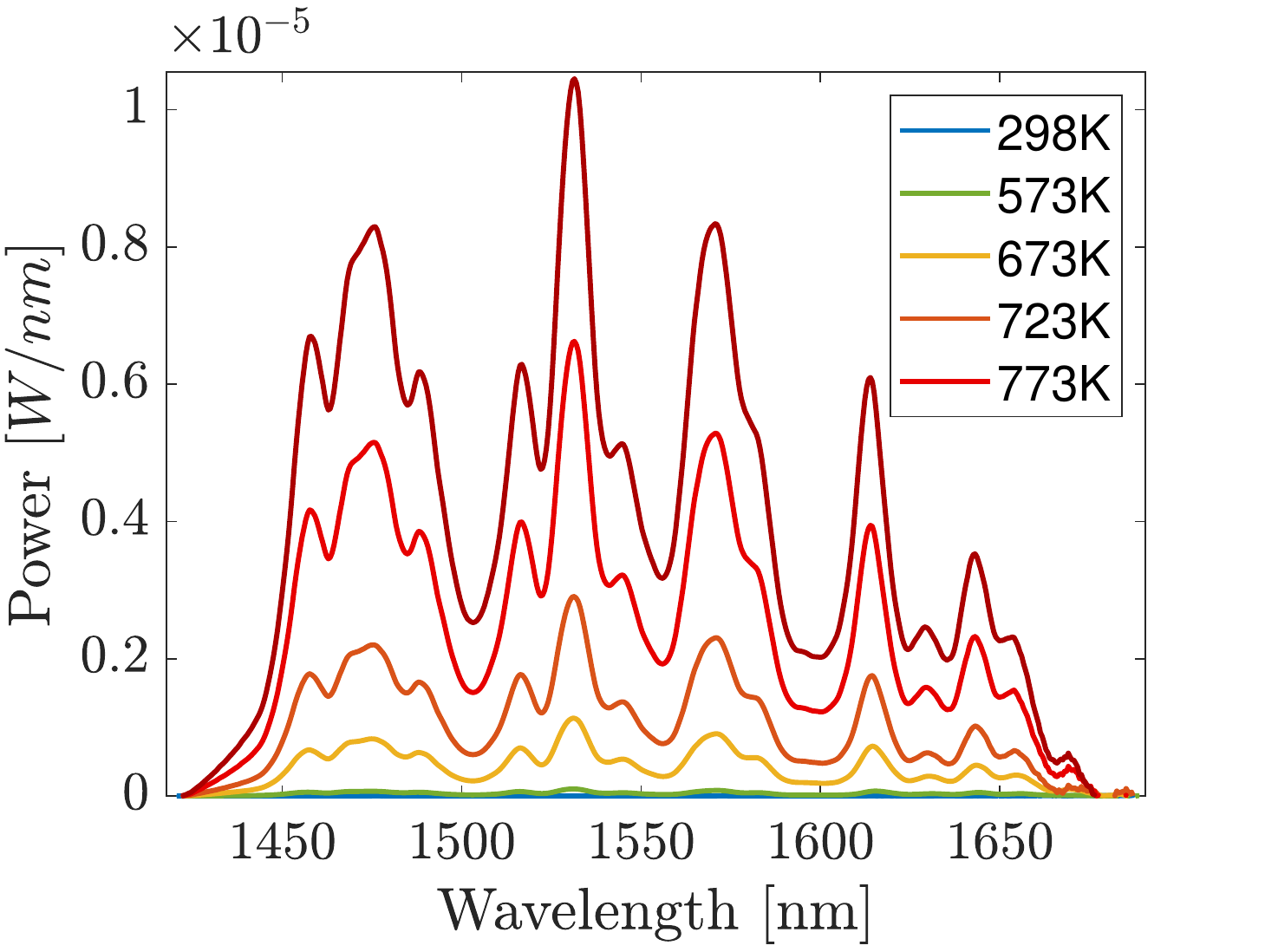}
         \caption{}
         \label{fig:Spectrum_Thermal_2}
     \end{subfigure}
     \hfill
     \begin{subfigure}[b]{0.3\textwidth}
         \includegraphics[width=\textwidth]{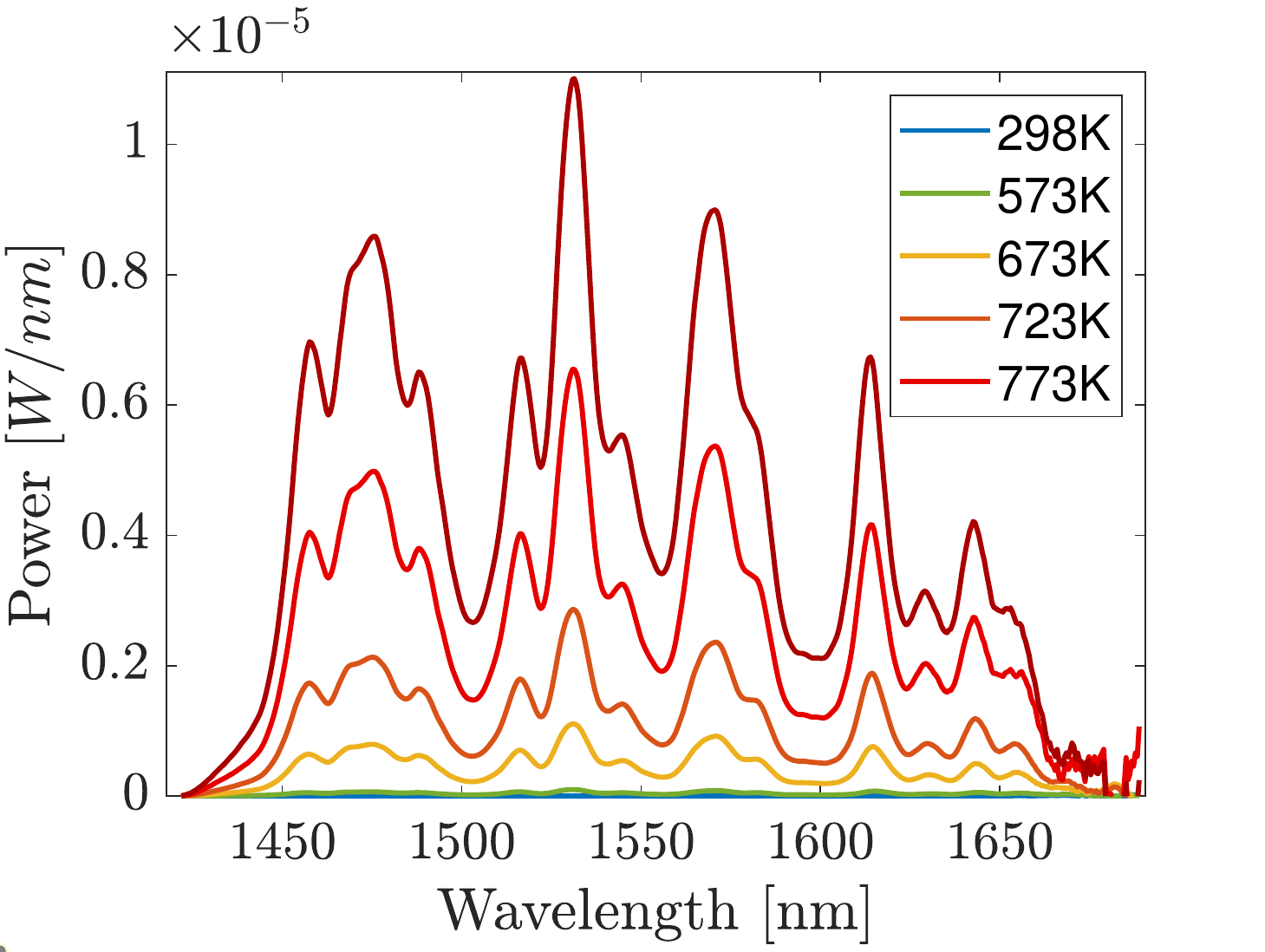}
         \caption{}
     \end{subfigure}
        \caption{Power spectra of thermal emissions of three Er:YAG concentrations in YAG crystals (0.2\%, 2\% and 3\%).}
\label{spectrum_thermal}
\end{figure*}

See results in FIG.~\ref{spectrum_thermal}.

\section{Normalization of the measured thermal spectrum}
\begin{figure}[h!]
\includegraphics[width=0.4\linewidth]{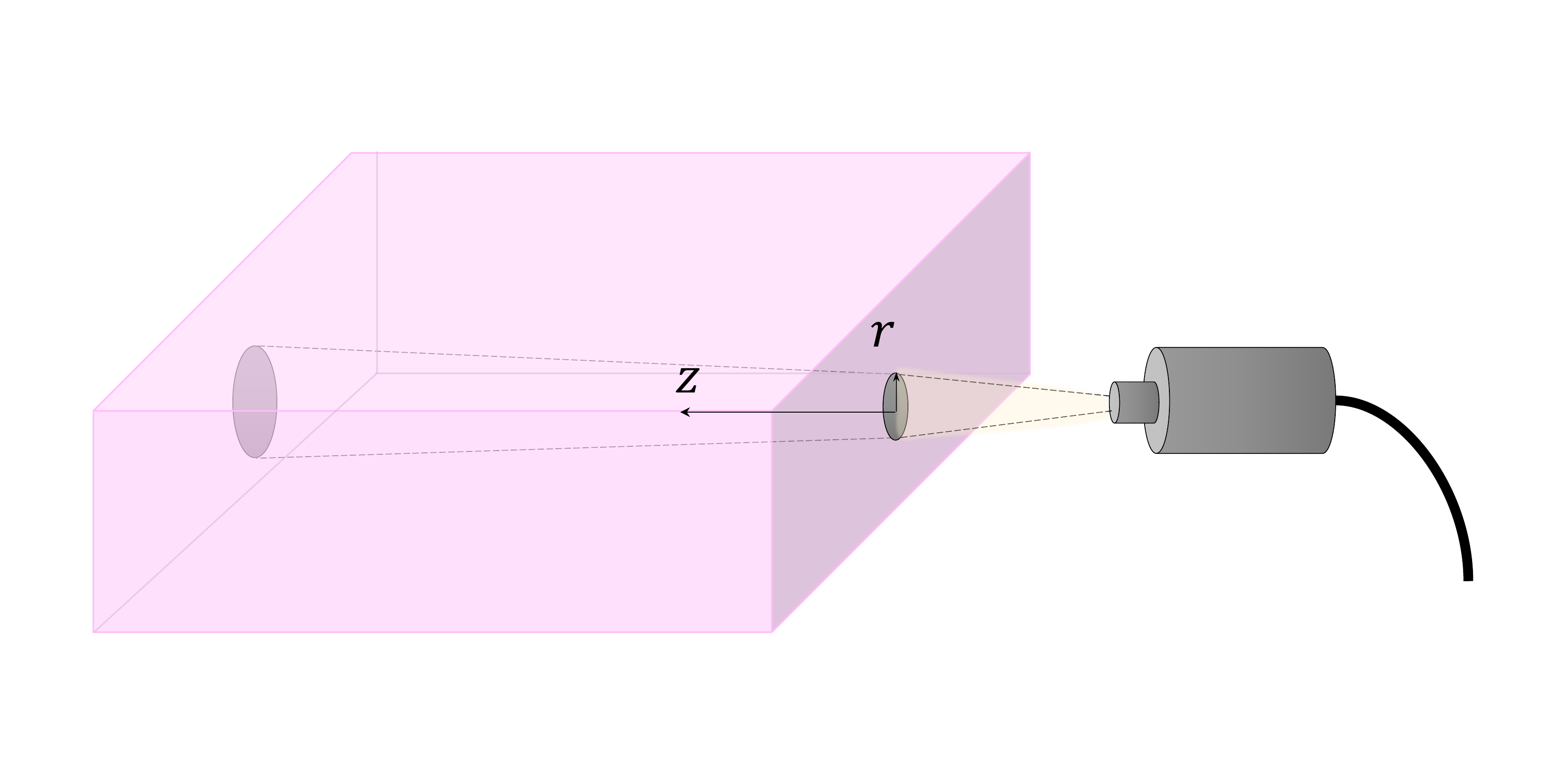}
\caption{Sample solid angle and emitting area.}
\label{fig_sample_solid}
\end{figure}

The experiment is performed in a free-space system, depicted in the paper. All samples are similarly aligned relative to the fiber. The numerical aperture and the distance at which the fiber is placed are such that the edges have no direct optical path to the fiber. Since Er:YAG has non-unity absorption at the measured spectra, we also account for the volume emission confined by the sample, 
presented in FIG.~\ref{fig_sample_solid}.
Therefore, we normalize the measured thermal emission power with the solid angle of the fiber and measured emitting volume.
\begin{equation}
    I_{emitted}[W] = I_{measured}[W] \cdot \frac{V_{sample}}{V_{cone}} \cdot \frac{4 \pi}{\Omega_{fiber}}
\end{equation}
Then $I_{emitted}$ is divided by the sample area and the $\pi$ solid angle to obtain the radiance:
\begin{equation}
\begin{aligned}
    R &&= &\frac{I_{emitted}}{\pi A_{sample}}
    \\
    &&= &\frac{I_{measured}}{\pi A_{sample}} \cdot \frac{V_{sample}}{V_{cone}} \cdot \frac{4 \pi}{\Omega_{fiber}}
    \label{eq. radiance}
    \\
    &&= &\frac{I_{measured} \cdot L}{\Omega_{fiber} \cdot V_{cone}}
    \left[ \frac{W}{sr \cdot m^2} \right]
\end{aligned}
\end{equation}
Here $V_{sample} = a \cdot L^2$ and $A_{sample}=4aL$ (upper and lower faces are reflecting), where $a$ is the sample's height and $L$ is the sample's length and width. In this case, the emissivity $\epsilon = \frac{R}{B.B}$ where $B.B$ denotes the black-body emission.
\section{Absorptivity measurement and normalization}
\begin{figure*}[h]
     \centering
     \begin{subfigure}{0.4\textwidth}
        \includegraphics[width=\textwidth]{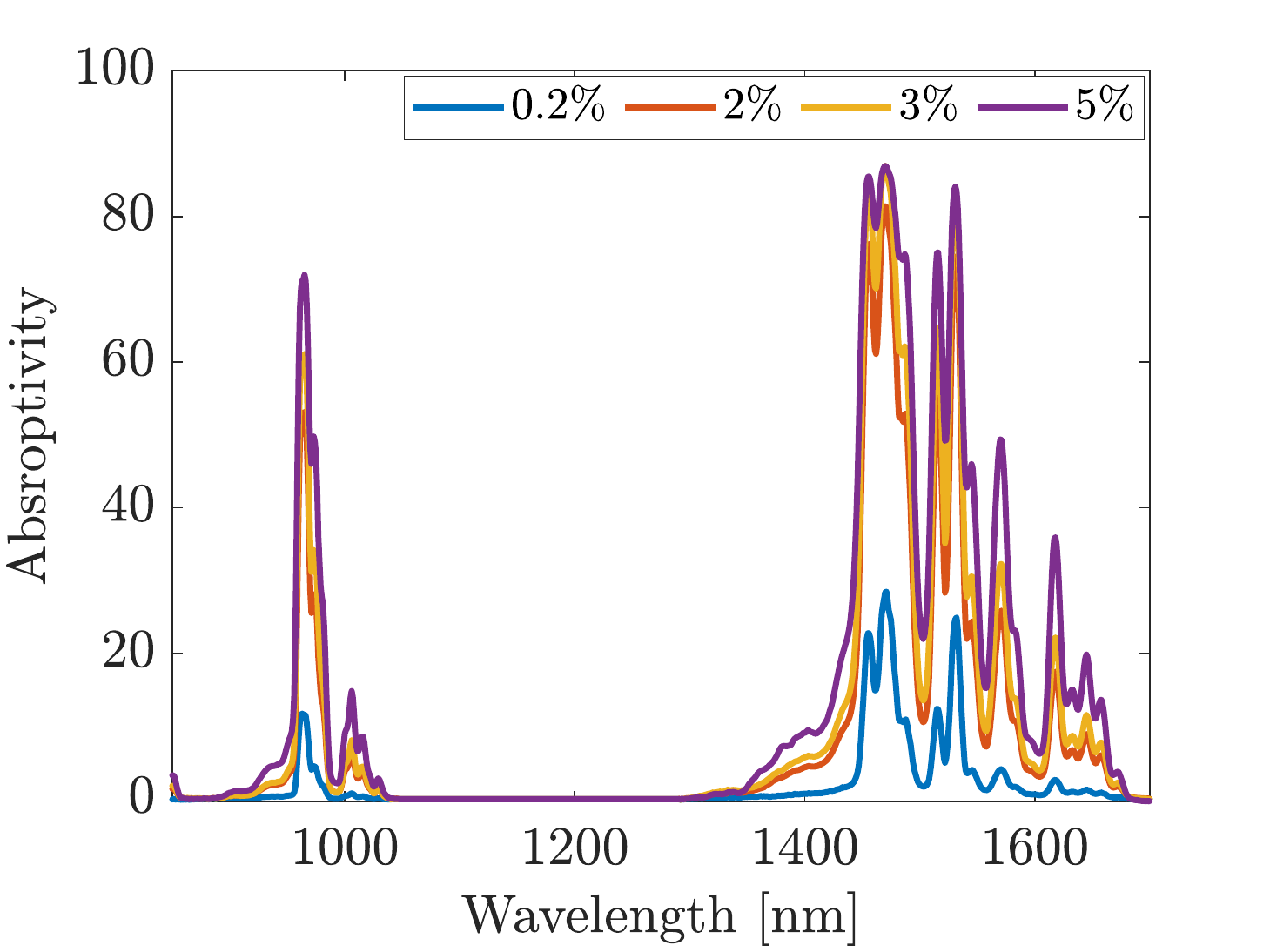}
         \caption{}
         \label{fig: abs vs WL}
     \end{subfigure}       
     \hfill
     \begin{subfigure}{0.4\textwidth}
       \includegraphics[width=\textwidth]{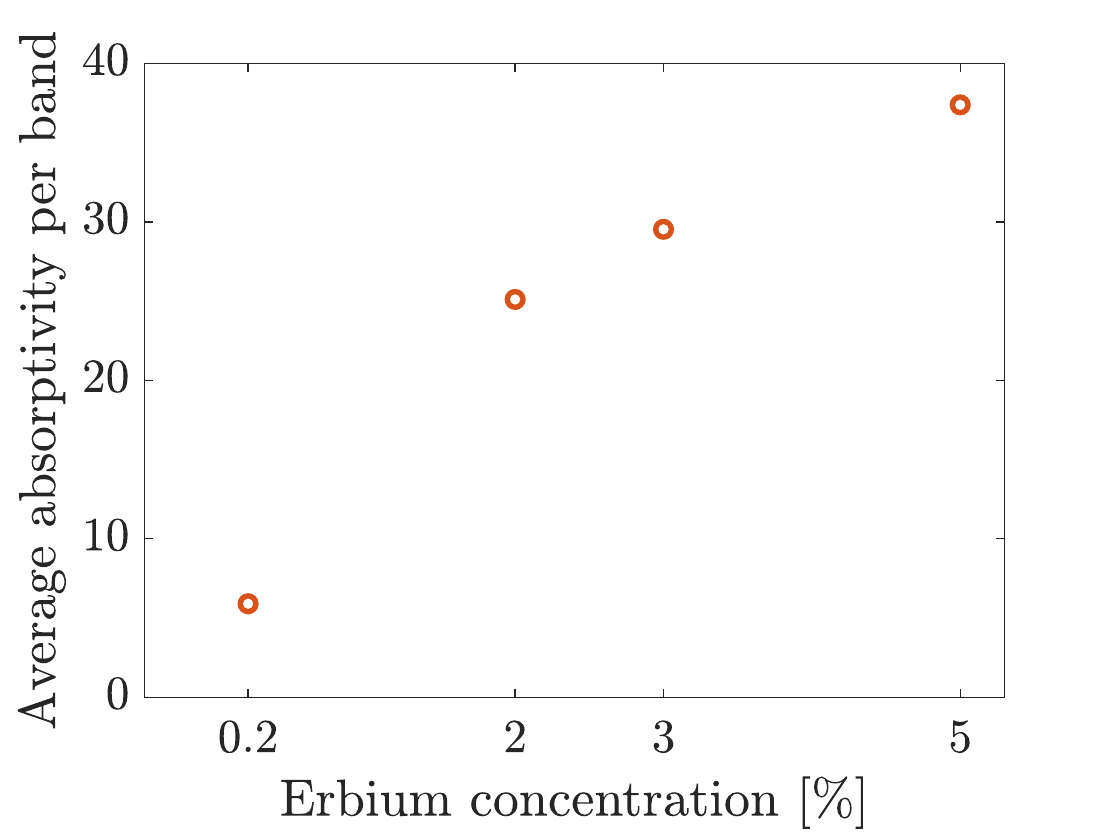}
         \caption{}
         \label{fig: avg abs vs concentration}
     \end{subfigure}
        \caption{Absorptivity of all Er:YAG samples. (a) The measured spectral absorptivity of all Er:YAG crystals. (b) The averaged absorptivity spectra, which are presented in (a), per band: 1450-1650 nm of all Er concentration samples.}
        \label{fig: Absorptivity}
\end{figure*}

In FIG. 4 in the paper, we present the absorptivity averaged by the wavelength band in the range of 1450-1650 nm, measured and normalized inside an integrating sphere. For this, we first measure the directional absorptivity spectra of all the Er:YAG samples in the range of 850-1680 nm at room temperature (Cary 5000 UV-VIS), presented in FIG.~\ref{fig: abs vs WL}.

\begin{figure}[H]
\centering
\includegraphics[width=0.4\linewidth]{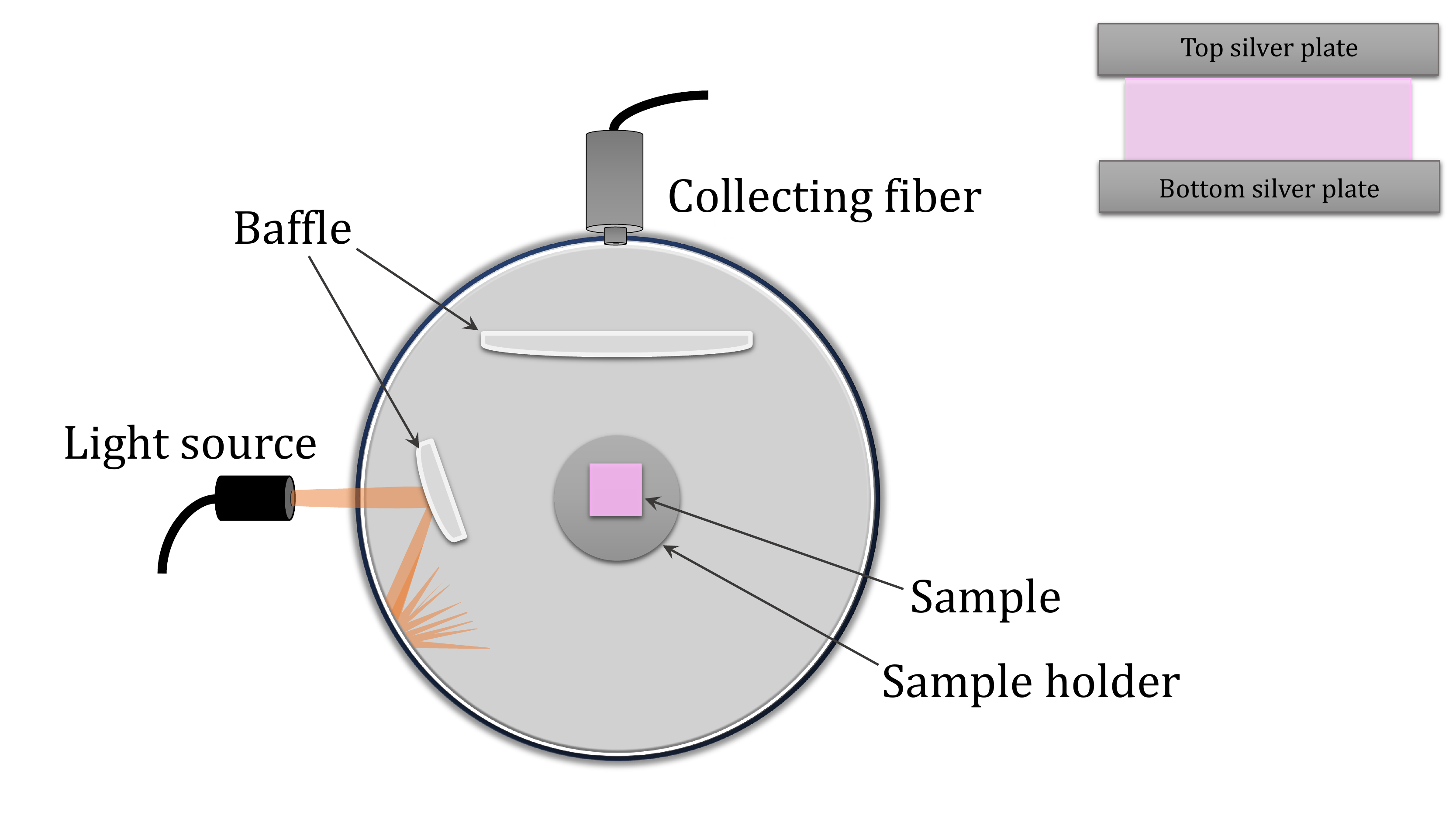}
\caption{\label{fig int sph} Isotropic absorptivity measurement system. (a) presents schematically the
measurement of the unabsorbed diode, where (b) illustrates the experimental system
for measuring the diode. The placing of the sample between the bottom and top silver plates
is presented in the inset of (a).}
\end{figure}

We acquire the hemispherical absorptivity at the same spectral band by placing each sample with its silver cap inside an integrating sphere, as presented in FIG.~\ref{fig int sph}. To follow the experimental conditions of the emissivity experiment, where the sample radiates only in the peripherical directions, we place each sample on the silver plate holder, and cover it with the top silver plate, as presented by the inset in FIG.~\ref{fig int sph}. In this experiment, the sample is illuminated with the uniformly scattered laser diode, 1550 nm wavelength (Prefile, LDS 1550), from the walls of the integrating sphere, when the light beam hits first the baffle placed at the entrance of the sphere.
The isotropic absorptivity is calculated as:
\begin{equation}
    \alpha_{isotropic} \left( \lambda = 1550 [nm]\right) = \frac{I_{empty} - I_{sample\;in}}{I_{empty}}
\end{equation}
where $I_{empty}$ and $I_{sample\;in}$ calibrated intensities at the diode emission wavelength when the sphere is empty and when the sample is placed inside the sphere, respectively. The ratio between the isotropic and directional absorptivity is the normalization factor by which the directional absorptivity at the measured spectrum is normalized.

\section{Emissivity independent of temperature}
\begin{figure} [H]
\centering
\includegraphics[width=0.4\linewidth]{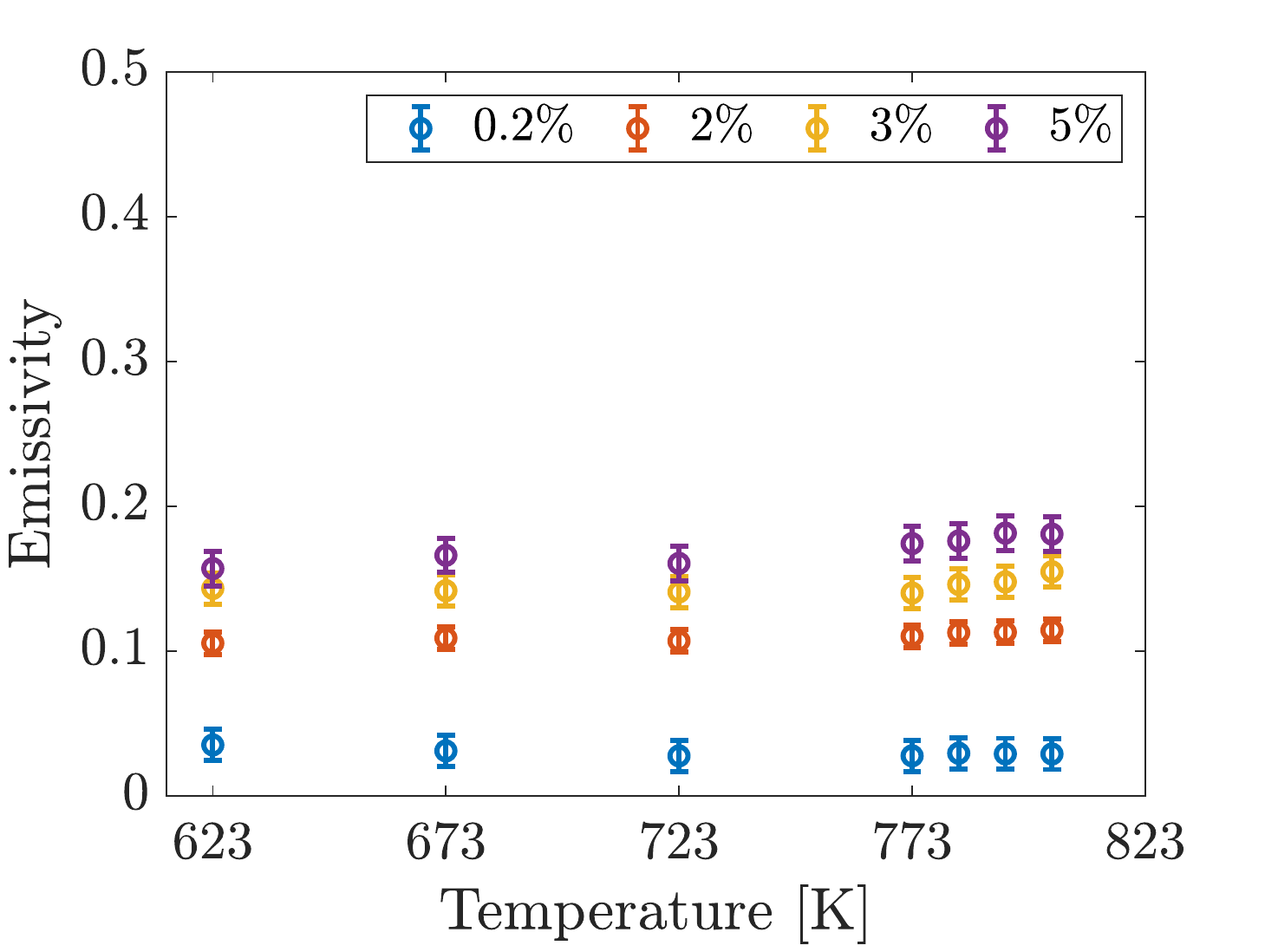}
\caption{\label{fig emissivity} Emissivity independent of temperature of all measured Er:YAG samples.}
\end{figure}

The generalized Kirchhoff’s law $\epsilon = \alpha \times (1-QE)$ is correct for any wavelength direction and $QE$.
For simplicity in the paper, we develop the temperature evolution of the emission for $\gamma_r$ and $\gamma_{nr}$ to be temperature independent.
We show that this assumption is true for the experiment with the Er:YAG samples.
Fig.~\ref{fig emissivity} depicts the emissivity values of all measured samples. These values are the spectral sum of the emissivity $\sum_{\lambda} \epsilon(\lambda)$, obtained as the ratio of the thermal and black-body emissions at the range of 1450-1650 nm. We can see that the emissivity is, to a good approximation, constant, and therefore, independent of temperature.
